\documentclass[12pt]{article}

\usepackage{mathrsfs}
\usepackage[T1]{fontenc}
\usepackage{mathpazo}
\usepackage{setspace}
\usepackage{amsfonts}
\usepackage{amssymb}
\usepackage{amsmath}
\usepackage{esint}                   
\usepackage{epsfig}
\usepackage{latexsym}
\usepackage{color}
\usepackage{graphicx}
\usepackage{hyperref}
\usepackage{nicefrac}
\usepackage[latin1]{inputenc}
\usepackage{pstricks}
\usepackage{slashed}
\usepackage{multirow}
\usepackage{ulem}
\usepackage{comment}
\usepackage[nosort]{cite}
\usepackage{datetime}
\usepackage{tikz-cd}
\usepackage{float}

\usepackage{comment}
\usepackage{ytableau}
\usepackage{braket}

 \usetikzlibrary{decorations.pathreplacing}

 \usetikzlibrary{decorations.pathreplacing}

\usepackage{pgfplots}
\usepackage{tikz}
\usetikzlibrary{arrows.meta}
\usepackage{caption}
\usepackage{subcaption}

\def\hybrid{\topmargin -30pt    \oddsidemargin 0pt 
        \headheight 0pt \headsep 0pt
        \textwidth 6.25in       
        \textheight 9.5in       
        \marginparwidth .875in
        \parskip 5pt plus 1pt   \jot = 1.5ex}

\hybrid

\def\baselinestretch{1.2}

\catcode`\@=11

\def\marginnote#1{}
\newcount\hour
\newcount\minute
\newtoks\amorpm
\hour=\time\divide\hour by60
\minute=\time{\multiply\hour by60 \global\advance\minute by-\hour}
\edef\standardtime{{\ifnum\hour<12 \global\amorpm={am}%
        \else\global\amorpm={pm}\advance\hour by-12 \fi
        \ifnum\hour=0 \hour=12 \fi
        \number\hour:\ifnum\minute<10 0\fi\number\minute\the\amorpm}}
\edef\militarytime{\number\hour:\ifnum\minute<10 0\fi\number\minute}

\def\draftlabel#1{{\@bsphack\if@filesw {\let\thepage\relax
   \xdef\@gtempa{\write\@auxout{\string
      \newlabel{#1}{{\@currentlabel}{\thepage}}}}}\@gtempa
   \if@nobreak \ifvmode\nobreak\fi\fi\fi\@esphack}
        \gdef\@eqnlabel{#1}}
\def\@eqnlabel{}
\def\@vacuum{}
\def\draftmarginnote#1{\marginpar{\raggedright\scriptsize\tt#1}}

\def\draft{\oddsidemargin -.5truein
        \def\@oddfoot{\sl preliminary draft \hfil
        \rm\thepage\hfil\sl\today\quad\militarytime}
        \let\@evenfoot\@oddfoot \overfullrule 3pt
        \let\label=\draftlabel
        \let\marginnote=\draftmarginnote
   \def\@eqnnum{(\theequation)\rlap{\kern\marginparsep\tt\@eqnlabel}%
\global\let\@eqnlabel\@vacuum}  }

\def\draft2{
        \def\@oddfoot{\sl preliminary draft \hfil
        \rm\thepage\hfil\sl\today\quad\militarytime}
        \let\@evenfoot\@oddfoot \overfullrule 3pt
        \let\marginnote=\draftmarginnote
   \def\@eqnnum{(\theequation)\rlap{\kern\marginparsep\tt\@eqnlabel}%
\global\let\@eqnlabel\@vacuum}  }

\def\preprint{\twocolumn\sloppy\flushbottom\parindent 2em
        \leftmargini 2em\leftmarginv .5em\leftmarginvi .5em
        \oddsidemargin -.5in    \evensidemargin -.5in
        \columnsep .4in \footheight 0pt
        \textwidth 10.in        \topmargin  -.4in
        \headheight 12pt \topskip .4in
        \textheight 6.9in \footskip 0pt
        \def\@oddhead{\thepage\hfil\addtocounter{page}{1}\thepage}
        \let\@evenhead\@oddhead \def\@oddfoot{} \def\@evenfoot{} }

\def\numberbysection{\@addtoreset{equation}{section}
        \def\theequation{\thesection.\arabic{equation}}}

\def\underline#1{\relax\ifmmode\@@underline#1\else
        $\@@underline{\hbox{#1}}$\relax\fi}

\def\titlepage{\@restonecolfalse\if@twocolumn\@restonecoltrue\onecolumn
     \else \newpage \fi \thispagestyle{empty}\c@page\z@
        \def\thefootnote{\fnsymbol{footnote}} }

\def\endtitlepage{\if@restonecol\twocolumn \else \newpage \fi
        \def\thefootnote{\arabic{footnote}}
        \setcounter{footnote}{0}}  

\catcode`@=12
\relax

\def\figcap{\section*{Figure Captions\markboth
        {FIGURECAPTIONS}{FIGURECAPTIONS}}\list
        {Figure \arabic{enumi}:\hfill}{\settowidth\labelwidth{Figure
999:}
        \leftmargin\labelwidth
        \advance\leftmargin\labelsep\usecounter{enumi}}}
 \relax
\def\tablecap{\section*{Table Captions\markboth
        {TABLECAPTIONS}{TABLECAPTIONS}}\list
        {Table \arabic{enumi}:\hfill}{\settowidth\labelwidth{Table
999:}
        \leftmargin\labelwidth
        \advance\leftmargin\labelsep\usecounter{enumi}}}
 \relax
\def\reflist{\section*{References\markboth
        {REFLIST}{REFLIST}}\list
        {[\arabic{enumi}]\hfill}{\settowidth\labelwidth{[999]}
        \leftmargin\labelwidth
        \advance\leftmargin\labelsep\usecounter{enumi}}}
 \relax
\makeatletter
\newcounter{pubctr}
\def\publist{\@ifnextchar[{\@publist}{\@@publist}}
\def\@publist[#1]{\list
        {[\arabic{pubctr}]\hfill}{\settowidth\labelwidth{[999]}
        \leftmargin\labelwidth
        \advance\leftmargin\labelsep
        \@nmbrlisttrue\def\@listctr{pubctr}
        \setcounter{pubctr}{#1}\addtocounter{pubctr}{-1}}}
\def\@@publist{\list
        {[\arabic{pubctr}]\hfill}{\settowidth\labelwidth{[999]}
        \leftmargin\labelwidth
        \advance\leftmargin\labelsep
        \@nmbrlisttrue\def\@listctr{pubctr}}}
 \relax
\makeatother

\def\be{\begin{equation}}
\def\ee{\end{equation}}
\def\ba{\begin{eqnarray}}
\def\ea{\end{eqnarray}}

\def\del{\partial}

\def\bx {{\bf x}}
\def\bw {{\bf w}}
\def\bX {{\bf X}}

\def\b{\beta}

\def\g{\gamma}

\def\D{\Delta}
\def\e{\epsilon}

\def\l{\lambda}
\def\L{\Lambda}
\def\s{\sigma}

\def\no{\noindent}

\def\qq{\qquad}

\def\IR{\relax{\rm I\kern-.18em R}}

\def\bx {{\bf x}}
\def\bw {{\bf w}}

\def\bv {{\bf v}}

\def\inv{^{\raise.0ex\hbox{${\scriptscriptstyle -}$}\kern-.05em 1}}

\def \ha {{\frac{1}{2}}}

\def \ov {\over}

\def\diag{{\rm diag}}

\def\half{{\textstyle {1 \over 2}}}

\newcommand{\bb}{\hskip -0.1cm}
\newcommand{\hb}{\hskip -0.05cm}

\def\kk{{\bf k}}
\def\rr{{\bf r}}
\def\ss{{\bf s}}
\def\zz{{\bf z}}

\def\bq{{\bf q}}

\usepackage{amsmath, amssymb, amsthm}
\usepackage{hyperref}
\usepackage{geometry}
\begin{document}


\renewcommand{\theequation}{\thesection.\arabic{equation}}
\csname @addtoreset\endcsname{equation}{section}

\begin{titlepage}
\begin{center}

\renewcommand*{\thefootnote}{\arabic{footnote}}

\phantom{xx}
\vskip 0.5in

{\large {\bf Interacting multicomponent $SU(N)$ magnets}} 

\vskip 0.5in

 {\bf Alexios P. Polychronakos}$^1$, \hskip .15cm
{\bf Konstantinos Sfetsos}$^{2}$

\vskip 0.15in

${}^1\!$ Physics Department, the City College of New York, NY 10031, USA \\
The Graduate Center of the CUNY, New York, NY 10016, USA\\
{\footnotesize{\tt apolychronakos@ccny.cuny.edu}}; {\footnotesize{\tt apolychronakos@gc.cuny.edu}}

\vskip 0.15in

${}^2\!$
Department of Nuclear and Particle Physics, \\
Faculty of Physics, National and Kapodistrian University of Athens, \\
Athens 15784, Greece\\
{\footnotesize{\tt ksfetsos@phys.uoa.gr}}

\vskip .3in
\today

\vskip .2in

\end{center}

\vskip .2in

\centerline{\bf Abstract}

\no
We study magnets composed of molecules with atoms interacting via $SU(N)$-invariant two-body interactions. Atoms carry states
transforming in generically different irreducible representations of $SU(N)$, and the interaction strength between atoms depends
on the specific atoms involved.
We derive the thermodynamics of these magnets in the mean-field approach, and find that they can exhibit mixed ferro-antiferromagnetic behavior.
We present a full analysis of the simplest nontrivial case of molecules consisting of two atoms in the spin-$j$ 
representation of $SU(2)$, which generalizes conventional magnetic systems, and derive its phase structure and transitions.

\vskip .4in

\vfill

\end{titlepage}
\vfill
\eject



\def\baselinestretch{1.2}
\baselineskip 20 pt

\newcommand{\eqn}[1]{(\ref{#1})}

\tableofcontents

\section{Introduction}

Magnetism with $SU(N)$ degrees of freedom is enjoying a revival, mostly because of its appearance in systems of ultracold molecules
\cite{Dud,Ghu,Gor,Zha,Mag,Cap,Muk}, and has been theoretically studied in spin chains \cite{Aff,Pola} and
interacting molecules on lattice cites \cite{KT,BSL,RoLa,YSMOF,TK,Totsuka,TK2}, also in the presence of external $SU(N)$ magnetic fields \cite{DY,YM,HM}.

\no
In our recent work \cite{PSferro, PSlargeN, PSferroN, PShigher, PScubic, LPSadj} we have derived the thermodynamics of $SU(N)$ ferromagnets in the mean field
approach and explored their phase structure,
which proves extremely rich and qualitatively different from that of ordinary $SU(2)$ ferromagnets. 
The systems studied consisted of elementary magnets ("atoms") each carrying internal degrees of freedom interacting through two-body or higher-body interactions
invariant under global $SU(N)$ transformations of the molecules' degrees of freedom. Our study revealed the existence of several critical temperatures
generalizing the Curie
temperature for $SU(2)$, metastability, hysteresis, and spontaneous breaking of the $SU(N)$ symmetry and of discrete symmetries
into various channels. In the case of two-body interactions the only relevant parameter is the rank of the group $N$,
while the presence of three-body interactions introduces an additional relevant parameter
that drives zero-temperature
(quantum) phase transitions. As the rank of the group $N$ becomes large, new critical phenomena emerge involving a triple critical point.

\no
The $SU(N)$ representation carried by each atom is a defining element of the ferromagnet. The original investigations were for the fundamental representation
\cite{PSferro, PSferroN, PScubic}.
In \cite{PShigher, LPSadj} we
developed the formalism for dealing with a general atom representation and studied in detail the cases of the symmetric, antisymmetric,
and adjoint irreducible representations, uncovering additional spontaneous symmetry breaking patterns and phases.

\no
In the present work we extend the analysis to magnets with composite atoms ("molecules") consisting of several components ("atoms"), each carrying a representation of $SU(N)$
and interacting through two-body terms. Each atom of a molecule can interact with the atoms of another molecule, leading to a matrix of coupling
constants. Such magnets can manifest ferromagnetic, antiferromagnetic, or mixed properties, depending on the structure of the coupling matrix.
The thermodynamic equilibrium equations and order parameters of these magnets are substantially more complicated than the ones for a single component, and their study
becomes mathematically challenging, requiring a combination of exact and numerical approaches. In this paper we derive the equilibrium equations of these magnets,
examine their general properties, and work out explicitly the simplest nontrivial case of two components in the spin-$j$ representation of
$SU(2)$, which is a generalization of the ordinary $SU(2)$ magnet.

\no
The organization of the paper is as follows:
In section \ref{model} we introduce the model of magnets with molecules having a general number of atoms in (generically different)
irreducible representations of $SU(N)$ and derive the equilibrium equations and stability conditions. In section \ref{su2} we analyze the case
of two $SU(2)$ atoms and derive its phase structure. Finally, section \ref{conclusions} contains some concluding remarks. Appendices A through E contain
derivations and proofs relevant to the analysis of the model.

\section{The multicomponent $SU(N)$ ferromagnet}\label{model}
 
 \subsection{The two-body Hamiltonian}
 \label{tbH}
 
 We consider $n \gg 1$ molecules at fixed positions, each carrying a set of $D$ internal states and interacting with two-body interactions depending on their
 distance. We assume the interactions are invariant under some symmetries, implied by the physics of the situation, and this restricts the form of the interaction
 Hamiltonian, as we will specify.
 
 \no
 If all internal states in each molecule are equivalent, the Hamiltonian is invariant under a common change of basis for these states and the symmetry is a
 global $SU(D)$ rotation, with the $D$ states transforming in the fundamental representation.
 In this case the Hamiltonian is of the exchange form,
 swapping the states of any two interacting molecules. There are, however, instances in which the symmetry is reduced. One such case involves states that
 transform under a higher, $D$-dimensional representation of a unitary group $SU(N)$, with $N<D$.
 Such cases,  involving the fundamental representation or higher ones, have been examined in our previous publications \cite{PSferro, PSlargeN, PSferroN, PShigher, PScubic, LPSadj}.
 
 \no
 A more general case is the one in which each molecule has several internal degrees of freedom ("atoms"), the states in each of them transforming under some
 representation of a
 global $SU(N)$ symmetry. The full Hilbert space of the molecule is the direct product of the Hilbert spaces of the atoms. Thus, the molecule can be viewed as consisting
 of several components (atoms), each with its own states. The full system, however, is fully coupled, as states in different atoms can interact.
 
 \no
A physically motivated construction of a Hamiltonian of the above type is given in Appendix \ref{microhami}. Here we present the general $SU(N)$-invariant model.
It involves molecules with $m$ atoms, the states in each atom transforming under a global $SU(N)$ group in (generally different)
irreducible representations (irreps) $r_s$, with $s=1,\dots,m$ labeling the atoms. The Hilbert space of each molecule is the direct product of the
Hilbert spaces for each atom $s$,
\be
{\cal H}_i = \prod_{s=1}^m \otimes {\cal H}_{s} \ ,\qq \dim({\cal H}_i) = \prod_{s=1}^m \dim(r_s)
\ee
and the full Hilbert space of the magnet is
\be
{\cal H} = \prod_{i=1}^n \prod_{s=1}^m \otimes {\cal H}_s = \Big(\prod_{s=1}^m \otimes {\cal H}_s\Big)^n\ .
\ee
Denoting $j_{s,i}^a$ the $SU(N)$ generators in irrep $r_s$ acting on the degrees of freedom $s$ of molecule $i$, the most general $SU(N)$ invariant Hamiltonian
with two-body interactions is
\be
H = \sum_{i,j=1}^n H_{ij} = \ha \sum_{i,j=1}^n \sum_{s,t=1}^m c_{st,ij}\sum_a j_{i,s}^a\, j_{j,t}^a\ ,
\label{Hgeneral}
\ee
with $c_{st,ij}$ real coupling constants obeying
\be
c_{st,ij} = c_{ts,ji}\ ,\qq c_{st,ii} = 0\ .
\ee
the latter condition eliminating molecule self-interactions.

\subsection{The thermodynamic free energy}

In the thermodynamic limit of a large number of molecules $n$, a mean field analysis similar to that of \cite{PSferro, PShigher} shows that
the magnet can be described in terms of the global $SU(N)$ generators $J_s^a$ and effective coupling constants $C_{st}$
 \be
  \label{Jsc}
 \begin{split}
 & J_s^a = \sum_{i=1}^n j_{s,i}^a \ ,\qq 
 [J_s^a, J_t^b] = {\rm i} \,\delta_{st}\,  f^{abc} J_s^c \ ,
 \\
 & C_{st} =  - \sum_{j=1}^n  c_{st,ij}\  ,
 \end{split}
 \ee
where we have assumed  homogeneity of the magnet, so that the sum over molecules $j$ in \eqn{Jsc} is independent of the molecule $i$, and
$C_{st}$ are globally defined quantities. The negative sign is included to make the ferromagetic case correspond to mostly positive $C_{st}$.
The resulting mean-field Hamiltonian is
 \be
 H = - \sum_{s,t=1}^m {C_{st}\over 2n} \sum_{a=1}^{N^2-1} J_s^a J_t^a
\label{Hcoupled}\ee
 and we refer the reader to \cite{PSferro, PShigher} for the details of its derivation. This Hamiltonian commutes with the total global $SU(N)$ generators
 \be 
J^a = \sum_{s=1}^m J_s^a
 \ee
 and, in principle, can be diagonalized within the space of irreps of $J^a$ contained in the direct product of the full set of individual irreps
 $(\prod_s \otimes r_s)^n$.
 
 \no
 Note that in the special case that all irreps $r_s$ are the same and the coupling matrix $C_{st}$ is invariant under permutations of $s$, that is,
 \be
 C_{st} = A \delta_{s,t} + B \ ,
 \label{CS}
 \ee
 the magnet has an additional symmetry $S_m$ that permutes the Hilbert spaces of the irreps. Cases with a subset of the irreps $s$ being the same
 and the coupling matrix being invariant under their permutations would have a reduced permutations symmetry.

\no
The limit $n\gg 1$ is essentially a semiclassical limit. In that limit, the typical irreps of $J_s^a$ correspond to large values of $SU(N)$ Casimirs
and the matrix elements of $J^a$ are of order $n$. This means that the $J^a$ become semiclassical
(in analogy to an irrep of $SU(2)$ of large spin $j$ becoming essentially a classical 3-vector of length $j$). To make this more explicit,
define operators ${\hat J}_s^a$ that have matrix elements of order one
\be
{\hat J}_s^a = {1\over n} J_s^a\ ,
\ee
which in the large-$n$ limit essentially commute. Hence, an eigenstate $|\l \rangle $  diagonalizes all ${\hat J}_s^a$'s with eigenvalues of 
order one. This means that terms quadratic in the currents of the type
$\displaystyle \sum_a {\hat J}_s^a {\hat J}_t^a$ commute in the large-$n$ limit and can all be evaluated using a common basis.

\no
Determining the values of terms $\displaystyle \sum_a {J}_s^a {J}_t^a=n^2\sum_a {\hat J}_s^a {\hat J}_t^a$  appearing in \eqn{Hcoupled},
requires expressing them in terms of the Casimirs of specific large irreps
of each $J_s^a$. Another crucial element that determines the entropy and thermodynamics of the system is the number
of states corresponding to each irrep of $J_s^a$, that is, the number of times each irrep of $J_s^a$ appears in the direct product of $n$ irreps $r_s$.
The above have been worked out in \cite{PScompo}, and we state the results below.

 \no
 Irreps of $SU(N)$ can be represented in terms of $N$ decreasing nonnegative integers $k_i$, $i=1,2,\dots,N$, denoted collectively as $\kk$
 and related to Young tableaux row lengths as
 $\ell_i = k_i -k_N +i-N$ \cite{PScompo}.
 The $k_i$ are, in principle, determined up to an overall additive constant (corresponding to the $U(1)$ charge of
 the $U(N)$ group). For an irrep arising in the decomposition of the direct product of $n$ irreps $r$, it is convenient to take $k_i$ to satisfy
 \be
 \sum_{i=1}^N k_i = n n_r +{N(N-1)\over 2}\ ,
 \ee
 with $n_r$ the number of boxes in the Young tablau corresponding to the irrep $r$. The quadratic Casimir of irrep $\kk$ has the form
 \be
 C_r^{(2)} = \half \sum_{i=1}^N k_i^2 -{\big ( n n_r + N(N-1)/2\big)^2 \over 2N} - {N(N^2-1)\over 24}\ .
 \label{cas2}\ee
 Apart from trivial constants, it is half the sum of squares of $k_i$, $\displaystyle \half\kk^2 = \half \sum_i k_i^2$.
 
 \no
 The multiplicity $d_{n,r;{\bf k}}$   of irrep $\kk$ in the decomposition of the direct product of $n$ irreps $r$ of $SU(N)$
 with character $\chi_r({\bf z})= {\rm tr}_r (\diag\{z_1,\dots,z_N\})$ is obtained through the expression 
 \be
 \sum_{\bf k} d_{n,r;{\bf k}}\, z_1^{k_1}\dots z_N^{k_N} = \D({\bf z}) \chi_r^n({\bf z})\ ,
 \ee
with $z_i$ auxiliary variables corresponding to the eigenvalues of the fundamental group element $U = \diag\{{\bf z}\}$,
and $\D({\bf z})$ the Vandermonde determinant
\be
\label{vanderm}
\Delta ({\bf z}) = \left|
\begin{array}{ccccccccc}
z_1^{N-1} & z_1^{N-2} & \cdots & z_1  & 1 \\
z_2^{N-1} & z_2^{N-2} & \cdots & z_2  & 1 \\
\vdots & \vdots & \ddots & \vdots & \vdots  \\
z_N^{N-1} & z_N^{N-2} & \cdots & z_N  & 1 \\
\end{array}
\right| \ .
\ee
Then $d_{n,r;{\bf k}}$ is obtained from a multiple integration in the complex $\bf z$-plane around the origin as \cite{PShigher}
 \be
 d_{n,r;{\bf k}} = {1\ov (2 \pi i)^N} \prod_{i=1}^N   \oint_{_C}  {d z_i\ov z_i^{k_i+1}}\, \D({\bf z})\, \chi_r^n({\bf z})\ .
 \label{dnrk}\ee

\no
It remains to identify the states that diagonalize the operators $J_s^a J_t^a$. The semiclassicality
argument of the previous paragraphs implies that, for $J_s^a$ and $J_t^a$ in irreps $\kk_s$ and $\kk_t$ respectively,
and inside sums involving $\kk_s$ and $\kk_t$,
\be
\sum_a J_s^a J_t^a = \sum_{i=1}^N k_{si} k_{ti} = \kk_s \cdot \kk_t\ ,
\ee
for large $n$. To justify this, we write these operators as
\be
\label{jjjk}
\sum_a {J}_s^a {J}_t^a = \half \sum_a ({J}_s^a + {J}_t^a)^2  - \half \sum_a ({J}_s^a)^2 - \half \sum_a ({J}_t^a)^2\ .
\ee
The last two terms are diagonal and equal to the quadratic Casimirs of $\kk_s$ and $\kk_t$. The states within irreps $\kk_s$ and
$\kk_t$ that realize irreps $\kk$ of $J_s^a + J_t^a$ diagonalize the full expression. A crucial set of facts is that the irreps $\kk$ of $J_s^a + J_t^a$
in the decomposition of $\kk_s \times \kk_t$ dominating in the thermodynamic sum over $\kk_s$, $\kk_t$, and $\kk$, are those with $\kk = \kk_s + \kk_t$,
and the number of these states in the sum is the number of irreps $\kk_s$ times the number of irreps $\kk_t$, to leading order in $n$. We refer to the
appendix \ref{partwo} for a proof of these nontrivial facts. Thus, ignoring the constants in \eqn{cas2},
\be
\sum_a J_s^a J_t^a \simeq \half (\kk_s + \kk_t)^2 -\half \kk_s^2 - \half \kk_t^2 = \kk_s \cdot \kk_t\ .
\ee
The effect of large $n$ is that, although in principle the states
diagonalizing $J_s^a + J_t^a$ and the states diagonalizing $J_s^a + J_t'^a$ are not the same, in the large-$n$ limit they
differ little and are, to leading order in $n$, all the states in $\kk_s \times \kk_t \times \kk_t'$. Further, the number of states (dimension) in each irrep $\kk_s$
is subleading compared to the multiplicity $d_{n,r_s;\kk_s}$ of the irrep $\kk_s$ in the Hilbert space. Putting everything together, we obtain for the partition
function $Z$ of the magnet in inverse temperature $\beta$
\be
Z = \sum_\text{states} e^{-\beta H} \simeq \prod_{s=1}^m \sum_{\kk_s} d_{n,r_s;\kk_s}\, e^{-\ha\beta \sum_{t,q=1}^m C_{tq} \kk_t \cdot \kk_q}\ .
\ee
The $\kk_s$ are of order $n$. Setting $\kk_s = n {\bf x}_s$, turning the sums over $k_{s,i}$ to integrals over $x_{s,i}$,
and taking the large-$n$ limit in the expression \eqn{dnrk} for the $d_{n,r_s;\kk_s}$, we finally obtain to leading order in $n$
\be
Z = \int \prod_{s=1}^m d{\bf x}_s\, d{\bf w}_s  \, e^{-n \beta F(\kk,\bw)}\ ,
\label{LZ}\ee
with
\be
F(\bx,\bw) = T\sum_s {\bx}_s \bb\cdot\bb{\bf w}_s - {1\over 2} \sum_{s,t} C_{st} {\bf x}_s \bb\cdot\bb {\bf x}_t
-T \sum_s \l_s ({\bw}_s)
\label{Fn}\ee
where
\be
z_{s,i} = e^{w_{s,i}}~~\text{and} ~~\l_s = \ln \chi_s\ .
\label{zwlx}\ee
The ${\bf x}_s$ are the order parameters of the magnet, $\bw_s$ are auxiliary dual variables that will be eliminated through their saddle point equations, and $F$ is
the free energy per molecule.
 
 \no
 The above derivation may appear somewhat heuristic. To fully prove its validity,
 we present in Appendix B the exact treatment of the case $m=2$ without any semiclassical approximations, which will justify the result.

 \subsection{Thermodynamic equilibrium equations}
 
 From the free energy $F(\kk,\bw)$ we can obtain the equilibrium equations and free energy $F(\bx)$. For large $n$, the exponent in \eqn{LZ}
 is of order $n$ and the integrals can be evaluated by the saddle point method. The corresponding saddle point equations are
 \be
 \begin{split}
 & {\partial F \over \partial x_{s,i}} = 0 : \qq T\, {w}_{s,i} = \sum_{t=1}^m C_{st}\, {x}_{t,i}  \ ,
 \\
 & {\partial F \over \partial w_{s,i}} = 0 :\qq  x_{s,i} = {\partial \l_s (\bw_s) \over \partial w_{s,i}} \ .
 \label{saddle}
 \end{split}
 \ee
Defining the $N$-vector function 
\be   
{\bf B}_s = \{ B_{s,i} \}\ ,\qq B_{s,i} (\bw_s )= \partial_{w_{s,i}} \l_s (\bw_s)\ ,
\ee
the saddle point equations can be written in the more compact vector notation
 \be
T {\bf w}_s = \sum_{t=1}^m C_{st} {\bf x}_t \ ,\qq  \bx_s = {\bf B}_s (\bw_s)\ .
\ee
 Solving the first equations for ${\bf w}_s$ gives the equilibrium equations for $\bx_s$
\be
\bx_s = {\bf B}_s \Big(\b\sum_{t=1}^m C_{st} {\bf x}_t \Big)\ .
\label{sa}
\ee
Plugging the solution for $\bw_s$ arising from the first of \eqn{saddle} back in the free energy yields the equilibrium free energy in \eqn{Fn}
\be
F_\text{eq} (\bx) = {1\over 2} \sum_{s,t} C_{st}\, {\bf x}_s \cdot{\bf x}_t
-T \sum_s \l_s \Big(\sum_t C_{st} \bx_t /T\Big) \ .
\label{Fe}\ee

\subsection{Constraints and stability}

The characters $\chi_s (\zz_s)$ are homogeneous polynomials in $z_{s,i}$ of degree $n_s$, the number of boxes in the
Young tableau of $r_s$. Summing the $w_{s,i}$ saddle equations in \eqn{saddle} over $i$ gives the constraints
\be
\sum_{i=1}^N x_{s,i} = n_s\ .
\label{cons}\ee
This means that solving these equations for the variables $w_{s,i}$, which is the proper procedure to eliminate
them, is possible only if the constraints are satisfied. Further, $\bw_{s,i}$ are only determined up to an overall additive constant, since 
the equilibrium equation for $w_{s,i}$ is invariant under such constant shifts.

\no
The derivation of the equilibrium free energy \eqn{Fe} followed the path of solving the saddle equations of $x_{s,i}$ for $w_{s,i}$. The off-equilibrium
free energy that should be used for deriving the stability of equilibrium configurations, however, is the one in which $\bw_s$ are eliminated using
their own saddle point equations.

\no
Calling $\bw_s (\bx_s)$ the solution of the second of \eqn{saddle}, with $\bx_s$ satisfying the constraints \eqn{cons},
the proper free energy valid for off-equilibrium configurations is
\be
F (\bx) = T\sum_s {\bf x}_s \bb\cdot\bb\bw_s(\bx_s) - {1\over 2} \sum_{s,t} C_{st} {\bf x}_s \bb\cdot\bb {\bf x}_t
-T \sum_s \l_s (\bw_s(\bx_s))
\label{Ft}
\ee
In the above, $\bx_s$ should satisfy the constraints \eqn{cons}, and in this case $F$ is invariant under shifts
$w_{s,i}(\bx_s) \to w_{s,i} (\bx_s) +c_s$. Further, variations of $\bx_s$ should preserve the constraints, that is, $\sum_i \delta x_{s,i} =0$.
In Appendix D we elaborate on the connection between the on- and off-equilibrium
expressions of the free energy \eqn{Fe} and \eqn{Ft}.

\no
The stability of equilibrium configurations is determined by the Hessian of the free energy \eqn{Ft} under variations of $x_{s,i}$ satisfying the
constraint $\sum_i \delta x_{s,i} =0$. We relegate the details to Appendix C and simply state the result.
Define the $N$-dimensional matrix
\be
\label{defls}
{\Lambda}_{s,ij} = {\partial^2 \lambda_s ({\bf w}_s) \over \partial_{w_{s,i}} \partial_{w_{s,j}}} \ ,
\ee
whose elements, we note, depend on the irrep $r_s$.
Then, the Hessian on the constraint space is the ($m N$)-dimensional square matrix with elements given by
\be
{\tilde {H}}_{si,tj} = T({\Lambda}_s )_{ij}^{-1} \delta_{st} - \, {C}_{st}\, \delta_{ij}  \ .
\label{hese}
\ee
The above Hessian has $m$ singular modes corresponding to the vectors $(u_{s'})_{s,i} = \delta_{s,s'}$ for $s' = 1,\dots,m$, which express deviations from the
constraint space. Stability requires the remaining eigenvalues of ${\tilde {H}}$ to be non-negative. The matrices ${\Lambda}_{s,ij}$ are positive definite
(a proof is provided in appendix E). Multiplying ${\tilde H}$ by the positive definite matrix $\Lambda_s$, 
the stability criterion can be expressed as
\be
{\bf C \Lambda}  \leqslant T\,  \mathbb{1}\otimes \mathbb{1}\ , 
\label{staba}
\ee
meaning that all the eigenvalues of the matrix $({\bf C  \Lambda })_{si,tj} = \Lambda_{s,ij} C_{st}$ must be smaller than $T$.
We note that the matrix $C_{st}$ has no a priori positivity
properties. 

\no
In general, the eigenvalues of ${\tilde {H}}_{si,tj}$ cannot be expressed in terms
of the eigenvalues of matrices $\Lambda_{s,ij}$ and $C_{st}$, so stability becomes a nontrivial condition. The condition simplifies in the
case of a magnet with an additional $S_m$ permutation symmetry, in which all irreps $r_s$ are the same and $C_{st}$ is of the form \eqn{CS},
and in phases where this symmetry is unbroken. In such phases, $\bx_s$ is the same for all $s$,
and thus the matrices ${\Lambda}_s = \Lambda$ are independent of $s$.
Then $\tilde {H} = T\,  \mathbb{1} \otimes  \Lambda_s^{-1} -  C \otimes \mathbb{1}$, and the matrix ${\bf C \Lambda } = C\otimes \Lambda$ is a 
direct products of matrices and admits eigenvectors of the form $v_{s,i} = a_i b_s$, with $a_i$
an eigenvector of $\Lambda$ with eigenvalue $\omega$ and $b_s$ an eigenvector of $\bf C$ with eigenvalue $\gamma$, giving
\be
\sum_{j,t} ({\bf C \Lambda})_{si,tj} \,a_j\, b_t = \omega  \gamma\, a_i\, b_s\ .
\ee
Stability requires all eigenvalues $\omega \gamma$ to be smaller than $T$. Since, due to \eqn{CS}, $\g = A$ (multiplicity $N-1$)
 or $\g = A+NB$ (multiplicity $1$), we must have
\be
\omega_\text{max} \big(A + N \max(B,0)\big) \leqslant T \  ,
\ee
where $\omega_\text{max}$ is the largest eigenvalue of $\Lambda$. For non-$S_m$ invariant magnets, or for phases that break $S_m$, we must find
the eigenvalues of the full, coupled form \eqn{hese} of $\tilde H$.

\subsection{Decoupling } \label{decoup}

In special situations, the equilibrium equations decouple into smaller sets of equations. 
One such case is when the coupling matrix $C_{st}$ is block diagonal. Then the free energy \eqn{Fn} breaks into non-interacting parts,
one for each block, and so do the equilibrium equations \eqn{sa}.

\no
Another such situation is when the coupling matrix is singular. Then, decomposing $\bx_s$ into eigenvectors of $C_{st}$, the eigenvectors corresponding
to zero eigenvalues drop out from the equations of the remaining eigenvectors and become
"subordinate," their equilibrium values determined entirely in terms of the remaining variables.
Further, the stability matrix $\bf C  \Lambda$ has $N$ zero eigenvalues for each zero mode of $\bf C$ (in addition to the $m$ zero eigenvalues of the modes
normal to the constraints), corresponding to the eigenvectors
\be
v_{s,i} = v_s a_i \  , \qquad C_{st} v_t = 0\ ,
\ee
for any $a_i$, which automatically satisfy the stability condition \eqn{staba}. The remaining eigenvectors $u_{s,i}$ of $\bf C \Lambda$
will be orthogonal to $v_{s,i}$, which means
that $\sum_s u_{s,i} v_s =0$ for all $i$. Therefore, both the equilibrium equations and the stability conditions are restricted to the subspace of $\bx_s$
corresponding to the nonzero eigenvalues of $\bf C$.

\section{An application: two coupled subsystems}\label{su2}

The free energy for two coupled $SU(N)$ sub-systems is given by \eqn{freeF}, or equivalently the $m=2$ version of \eqn{Fn},
which, using the notation for the coupling matrix $C_{st}$
\be
\label{cccpl}
(C_{st}) =   \left(
\begin{array}{cc}
c_a & c \\
c & c_b
\end{array}
\right)\ ,
\ee
 we rewrite here explicitly as\footnote{For later convenience we denote $s=1$ as $a$ and $s=2$ as $b$. Since we will eventually specialize to the $SU(2)$ case, 
 there should be no confusion with the $SU(N)$ group indices $a,b$.}
\be
\label{jjje}
\begin{split}
F({\bf x},{\bf w}) = &  T \sum_{i=1}^N (x_{a,i} w_{b,i} + x_{b,i} w_{b,i}) - {c_a\ov 2} \sum_{i=1}^N x_{a,i}^2 -  {c_b\ov 2} \sum_{i=1}^N x_{b,i}^2
- c \sum_{i=1}^N x_{a,i}  x_{b,i}
\\
& - T \ln \chi_a (w_a)  - T \ln \chi_b (w_b) \ ,
\end{split}
\ee
where $\chi_a$ and $\chi_b$ are the characters of the two irreps, generally different. The constraints 
\be
\sum_{i=1}^N x_{a,i}  = n_{a}\ ,\qq \sum_{i=1}^N x_{b,i}  = n_{b}\ ,
\ee
are also obeyed, with $n_{a}$ and $n_b$ being the number of boxes in the Young tableau for each representation. 

\no
To proceed we need an explicit expression of the characters in terms of the $w$-parameters. 
We utilize the relation between states in the fermionic representation and the character given by \cite{PScompo}
\be
\label{psidchi}
\psi_{\bf k} ({\bf z}) = \D({\bf z}) \chi_{\bf k}({\bf z})\ ,
\ee
The fermionic wavefunction corresponding to a general representation is given by the Slater determinant
\be
\psi_{\bf k} ({\bf z}) = \left|
\begin{array}{ccccccccc}
z_1^{k_1} & z_1^{k_2} & \cdots & z_1^{k_{N-1}}  & z_1^{k_N} \\
z_2^{k_1} & z_2^{k_2} & \cdots & z_2^{k_{N-1}}  & z_2^{k_N} \\
\vdots & \vdots & \ddots & \vdots & \vdots  \\
z_N^{k_1} & z_N^{k_2} & \cdots & z_N^{k_{N-1}} & z_N^{k_N} 
\end{array}
\right|\  ,
\label{slater}
\ee
where the Vandermonde determinant is given by \eqn{vanderm}
and $z_i = e^{w_i}$.
The normalizations are such that, for the singlet representation, $k_i= N-i$, $i=1,2,\dots ,N$ and the character is unity.
For higher irreps, the $k_i$ increase by the number of boxes in the corresponding row of the YT with length 
$\ell_i = k_i - k_N + i-N$, as previously mentioned.
Once the expressions for the characters are obtained, the saddle point equations for $w_i$ are obtained as
\be
\label{xaxb}
x_{a,i}= \del_{w_{a,i}} \ln \chi_{a}(w_a)\ ,\qq x_{a,i}= \del_{w_{b,i}} \ln \chi_{b}(w_b)\ ,
\ee
which determine the $w_i$ in terms of the $x_i$. The equilibrium equations for the $x_i$ read
\be
\label{hjh2}
T w_{a,i} - c_a x_{a,i} -c x_{b,i} = 0 \ ,\qq T w_{b,i} - c_b x_{b,i} -c x_{a,i} = 0 \ .
\ee

\subsection{The  $SU(2)$ case: Generalities}

As this is the simplest case we first examine our general set up for the one-atom case corresponding to $m=1$.
That can be obtained by setting the couplings $c_b$ and $c$ to zero. We also  drop the index $a$ in $c_a$ and $x_{a,i}$.
In this case, the spin-$j$ representation corresponds to $2 j$ boxes in the Young tableau. 
Taking $k_1=1+2 j$ and $k_2 =0$ in this irrep, the corresponding state is $\psi_j = z_1^{1+2 j}- z_2^{1+2 j}$ and the Vandermonde determinant is just
$\D=z_1-z_2$.
Hence, using \eqn{psidchi}, the character is 
\be
\label{xi1}
\chi_j = {z_1^{1+2 j}- z_2^{1+2 j}\ov z_1-z_2}= e^{j(w_1+w_2)} {\sinh {(j+1/2) (w_1-w_2)}\ov \sinh {1/2 (w_1-w_2)}}\ 
\ee
and we have the constraint  $x_1+x_2=2 j$. 
The $U(1)$ factor $e^{j(w_1+w_2)} $ is physically irrelevant and can be dropped. Hence, denoting $w=w_1-w_2$, the character becomes
\be
\label{xi2}
\chi_j =  {\sinh {(j+1/2) w}\ov \sinh {w/2}}\ .
\ee
Omitting the $U(1)$ factor effectively shifts $x_i\to x_i+j$, so the new $x_{1,2}$ sum to zero and take values in the range $[-j,j]$. We thus
define $x_1=j x$ and $x_2=-j x$ with $x\in [-1,1]$. In terms of $x$ and $w$, the free energy becomes
\be
F(x,w) = T j x w - c j^2  x^2 - T \ln {\sinh {(j+1/2) w}\ov \sinh {w/2}}\ .
\ee
The equilibrium equations for $x$ and $w$ are, respectively,
\be
 w = {2\ov T} j c x\ ,\qq   x = B_j(j w) \ ,
\ee
where 
\be
B_j(s) = {1+2 j\ov 2j} \coth  {1+2 j\ov 2j} s -{1\ov 2j} \coth {s\ov 2 j}\ ,
\ee
is the Brillouin function which is odd and monotonically increasing in $s$ from $-1$ to $1$. Eliminating $w$ we get the transcendental equation 
\be
\label{sssi}
 x = B_j(2 c j^2 x/T)\ .
\ee
The length of the Young tableau corresponding to the solution of the above equation is $\ell = 2 j x n$, with the number of molecules $n\gg 1$.
This is the standard result, obtainable also by elementary means in statistical mechanics in the mean field approximation. 
The solution to this equation is  $x=0$ for $T>T_c$ and $x\neq 0$ for $T<T_c$, where the critical temperature is
\be
T_c = \frac{2}{3} c\, j (j+1) \ .
\ee
Note that the classical case $j\to \infty$ is also attainable by first rescaling $c\to c/j^2$. Then 
\be
 x = L(2 c x/T)\ ,
\ee
where 
\be
L(s) = \coth  s -{1\ov s}\ ,
\ee
is the Langevin function, the large-spin limit of the Brillouin function. 

\no
Turning to the two-atom case with $m=2$, where all parameters $c_a, c_b$ and $c$ are in general nonzero, we parametrize, similar to the
single atom case,
\be
\label{fhhh}
x_{a,1}= -x_{a,2} = j_a x\ , \quad x_{b,1}= -x_{b,2} = j_b y \ ,\quad  w_a=w_{a,1}- w_{a,2} \ ,\quad w_b=w_{b,1}- w_{b,2}\,  .
\ee
Then, the free energy \eqn{jjje} becomes 
\be
\label{F22free}
\begin{split}
F(x,y,w_a,w_b) = &\, T j_a  w_a x +  T j_b  w_b y  - c_a j_a^2 x^2 - c_b j_b^2 y^2 -2 c j_a j_b x y 
\\
& 
- T \ln {\sinh {(j_a+1/2) w_a}\ov \sinh {w_a/2}} - T \ln {\sinh {(j_b+1/2) w_b}\ov \sinh {w_b/2}}\ .
\end{split}
\ee
The equilibrium equations with respect to $x$ and $y$ are
\be
\label{w1w2}
w_a = {2\ov T} (c_a j_a x + c j_b y)\ ,\qq  w_b = {2\ov T} (c_b j_b y + c j_a x)\ ,
\ee
and those with respect to $w_a$ and $w_b$ read
\be
\label{xyb}
 x = B_{j_a}(j_a w_a)\ ,\qq  y = B_{j_b}(j_b w_b)\ .
\ee
The total spin of the set of atoms $a$ and $b$ corresponding to solutions of \eqn{w1w2} and \eqn{xyb} are $J_a = n j_a |x|$ and $J_b = n j_b |y|$,
proportional to the number of molecules $n$.

\no
This system  is invariant under the transformation $(x,y)\to (-x,-y)$. In addition,
the case $c<0$ is recovered by flipping the sign of $y$. Hence, we may restrict to $x>0$ and $c>0$, whereas $c_a$, $c_b$, and $y$ 
can have any sign. Further, by rescaling the temperature $T$ we may set $c=1$.
The case $c=0$ corresponds to two uncoupled ordinary magnets.

\subsubsection{Stability analysis}

\no
The stability of the system can be assessed by examining the general condition \eqn{staba}. In our case the, matrix $\L$ is a four-dimensional 
square matrix with elements
\be
\L_{s,ij} = (\mathbb{1}- \s_1)_{ij}  \L_s\ ,\qq s=a,b\ ,\qq i,j=1,2\ ,
\ee
where $\s_1$ is the standard Pauli matrix and $\L_{a,b}$ are the positive quantities
\be
\L_a = j_a^2\, B'_{j_a}(j_a w_a)\ ,\qq  \L_b = j_b^2\, B'_{j_b}(j_b w_b)\ ,
\ee
with  $w_a$ and $w_b$ given by \eqn{xyb}.
The matrix elements of  ${\bf C\L}$ are
\be
({\bf C\L})_{st,tj} = \L_s  (\mathbb{1}- \s_1)_{ij} C_{st} \  ,
\ee
where $C_{st}$ are given by \eqn{cccpl}. This matrix is diagonalized by eigenvectors of the form $u_s v_i$.
The eigenvector 
$v_1 = v_2 =1$ corresponds to a zero eigenvalue for any ${\bf u}$ and hence it is doubly degenerate. 
The orthogonal vector 
$v_1 = -v_2 =1$ diagonalizes the $i,j$ matrix with eigenvalue 2, and \eqn{staba} reduces to the condition that the matrix 
\be
\label{Csta}
\left(
\begin{array}{cc}
T- 2 \L_a c_a &-2  \L_a c \\
-2 \L_b c &T -2  \L_b c_b
\end{array}
\right)
\ee
should have positive eigenvalues. Even though it is not symmetric, this matrix has real eigenvalues. Then the positivity criterion is satisfied by requiring positivity
of the determinant and of one of the diagonal entries of the matrix (positivity of the other diagonal entry is then assured). Explicitly, we require that
\be
\label{stabb4}
\begin{split}
& T^2 - 2 (\L_a c_a + \L_b c_b ) T + 4 \L_a \L_b  (c_a c_b -c^2)>0\  ,
\\
& T > 2 \L_a c_a\quad  ({\rm equivalently}\ T > 2 \L_b c_b )\ .
\end{split}
\ee
For the case $j_a=j_b=\ha$, where $B_{1/2}(s)=\tanh s$, $\L_{a,b}$ can be expressed in terms of $x,y$ as
\be
\label{stabb41}
\begin{split}
& x= \tanh {w_a\ov 2}\ ,\qq y= \tanh {w_b\ov 2}\ , 
\\
&  \L_a= {1\ov 4 \cosh^2 {w_a\ov 2}} = {1-x^2\ov 4}\ ,\qq \L_b= {1\ov 4 \cosh^2 {w_b\ov 2}} =  {1-y^2\ov 4}\ .
\end{split}
\ee

\subsection{The  $SU(2)$ case for spin 1/2}\label{su2h}

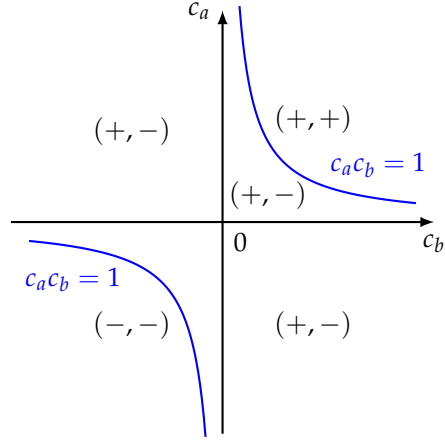
\begin{figure}[htbp]
    \centering
    \begin{tikzpicture}[>=latex, scale=.8]
        \draw[thick, ->] (-3.5, 0) -- (3.5, 0) node[below] {\footnotesize $c_b$};
        \draw[thick, ->] (0, -3.5) -- (0, 3.5) node[left] {\footnotesize $c_a$};
        \node[below right] at (0,0) {\footnotesize $0$};
        \draw[blue, thick, domain=0.28:3.2, samples=100] plot (\x, {1/\x});
        \draw[blue, thick, domain=-3.2:-0.28, samples=100] plot (\x, {1/\x});
        \node[blue, right] at (1.6, 0.95) {\footnotesize $c_a c_b = 1$};
        \node[blue, left]  at (-1.5, -0.95) {\footnotesize $c_a c_b = 1$};
        \node at (1.5, 1.7) {\footnotesize $(+,+)$};
        \node at (0.75, 0.45) {\footnotesize $(+,-)$};
        \node at (-1.5, 1.5) {\footnotesize $(+,-)$};
        \node at (1.5, -1.7) {\footnotesize $(+,-)$};
        \node at (-1.5, -1.7) {\footnotesize $(-,-)$};
    \end{tikzpicture}
    \vskip -.2 cm 
    \caption{\small{Signs of the parameters $(T_+,T_-)$ in the $(c_a\!-\!c_b)$-plane.}}
    \label{fig1}
\end{figure}

In the simplest case with $j_a=j_b=\ha$ the system of equilibrium equations becomes
\be
\label{xyhalf}
T \ln {1+x\ov 1-x} = c_a x + y\ ,\qq  T \ln {1+y\ov 1-y} = c_b y + x\ ,
\ee
where we used $\displaystyle \tanh^{-1} s= \ha \ln {1+s\ov 1-s}$, with $x,y\in [-1,1]$, and also set $c=1$ with no loss of generality, as argued after
\eqn{xyb}.
We may use the first equation to solve for $y=y(x)$ and substitute into the second one, obtaining a single transcendental equation for $x$
\be
\label{Txx}
\begin{split}
& T \bigg( \ln{1+y(x)\ov 1-y(x)} - c_b \ln {1+x\ov 1-x}\bigg) - (1-c_a c_b )x=0 \ ,
\\
& 
y(x) = T  \ln{1+x\ov 1-x} - c_a x\ ,
\end{split} 
\ee
We can see that the left hand side of \eqn{Txx} behaves as $4 (T-T_+)(T-T_-)x + {\cal O}(x^3)$ near $x=0$, where
\be
\label{tpmd}
T_\pm = {1\ov 4} \big(c_a+c_b \pm \sqrt{\D}\,\big)\ , \quad \D = (c_a-c_b)^2+4 \ ;\qq \bb  T_- < c_a, c_b < T_+\  .
\ee
Hence, the solutions depend on the magnitude of the temperature $T$ relative to the parameters $T_\pm$, which can have either sign.
Note that $T_\pm$ are the eigenvalues of the coupling matrix $\half {\bf C}$, with $\bf C$ as in 
\eqn{cccpl}. This is related to the fact that the stability matrix \eqn{Csta} has a zero eigenvalue for $T=T_\pm$ for the state $x=y=0$, in which case 
\eqn{stabb41} implies $\Lambda_a = \Lambda_b = {1\over 4}$. This translates to the vanishing of the derivative of the left hand side of the first of \eqn{Txx} at $x=0$.
The above facts play a role in the analysis of the solutions. The signs of $T_\pm$ are depicted in fig.~\ref{fig1}.

\no
We distinguish three cases in the $c_a, c_b$ parameter space:

\no
$\bullet$ \underline {$c_a, c_b>0,~ c_a c_b>1$:} Then both $T_\pm>0$.
If $T>T_+$ then the only solution is $x=0$ (so also $y=0$), whereas for temperatures in the range $T_-<T<T_+$ there is another solution as depicted in fig. \ref{fig2},
which is found numerically.


\pgfplotsset{compat=newest}

\begin{figure}[ht]
\centering

\begin{minipage}{0.48\textwidth}
\centering

\begin{tikzpicture}
\begin{axis}[
       width = 6.8cm,
        height = 5.5cm,
    axis lines=middle,
    axis line style={-{Latex[length=2.5mm,width=1.5mm]},thick},
    xmin=0, xmax=1.08,
    ymin=0, ymax=8.5,
    domain=0:0.96,
    samples=150,
    xlabel={},
    ylabel={},
    xlabel style={at={(rel axis cs:1.05, 0.0)}, anchor=west},
    xtick=\empty,
    ytick=\empty,
    enlargelimits=false,
    clip=false
]

\addplot[
    blue!70!black,
    thick
]
{x / (1 - 0.94*x*x)};

\node[black, below left] at (axis cs:0,0.9) {\footnotesize{$0$}};
  \node[black] at (1.05, .7) {\footnotesize $x$};
   \node[black] at (.6, 4.5) {\footnotesize $T_+<T$};

\end{axis}
\end{tikzpicture}

\end{minipage}
\hfill
\hskip -2 cm
\begin{minipage}{0.48\textwidth}
\centering

  \hskip -2 cm
\begin{tikzpicture}
\begin{axis}[
       width = 6.8cm,
        height = 5.5cm,
    axis lines=middle,
    axis line style={-{Latex[length=2.5mm,width=1.5mm]},thick},
    xmin=0, xmax=0.72,
    ymin=-0.75, ymax=0.75,
    domain=0:0.685,
    samples=150,
    xlabel={},
    ylabel={},
    xlabel style={at={(rel axis cs:1.05, 0.5)}, anchor=west},
    xtick=\empty,
    ytick=\empty,
    enlargelimits=false,
    clip=false
]

\addplot[
    blue!70!black,
    thick
]
{-1.8*x + 12*pow(x,5)};

\node[black, below left] at (axis cs:0,0.15) {\footnotesize{$0$}};
  \node[black] at (.7, .12) {\footnotesize $x$};
   \node[black] at (.4, .4) {\footnotesize $T_-<T<T_+$};

\end{axis}
\end{tikzpicture}

\end{minipage}

\vskip 0.2cm
\captionsetup{
    labelfont={small},
    textfont={small}
}
\caption{Plots of the left hand side of \eqn{Txx}  for $T_+<T$  (left) and for $T_-<T<T_+$ (right).
As $T\lesssim T_+$ the nontrivial solution $x\gtrsim 0$.}
\label{fig2}

\end{figure}
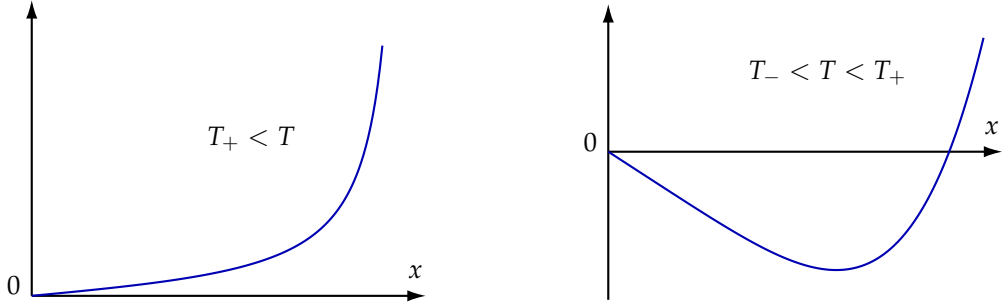

\no
If both $c_a , c_b >1$, there is a critical temperature $T_c <T_-$
such that, for $T<T_c$, a third solution appears, as depicted in fig.~\ref{fig3}.
The temperature $T_c$ is obtained numerically as a simultaneous solution of \eqn{Txx} and its first derivative with respect to $x$. 

\pgfplotsset{compat=newest}

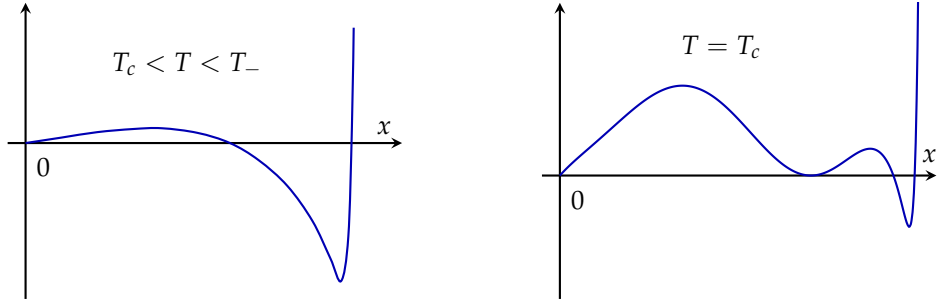
\begin{figure}[htbp]
\centering
    \begin{minipage}[b]{0.48\textwidth}
    \centering
    \begin{tikzpicture}
    \begin{axis}[
        axis lines = middle,
        axis line style = {->, >=stealth, thick},
        ticks = none,
        xlabel = {${}$},
        ylabel = {${}$},
        xlabel style = {anchor=north west},
        ylabel style = {anchor=south east},
        xmin = -0.05, xmax = 1.05,
        ymin = -2.0, ymax = 1.8,
        width = 6.8cm,
        height = 5.5cm
    ]
        \addplot [
            smooth,
            tension = 0.6,
            color = blue!70!black,
             thick
        ] coordinates {
            (0.00, 0.00)
            (0.10, 0.07)
            (0.20, 0.14)
            (0.30, 0.18)
            (0.38, 0.19)
            (0.48, 0.13)
            (0.57, 0.00)
            (0.65, -0.22)
            (0.73, -0.55)
            (0.80, -1.00)
            (0.85, -1.48)
            (0.88, -1.77)
            (0.90, -1.20)
            (0.91, 0.20)
            (0.915, 1.48)
        };
        
        \node[black, below left] at (axis cs:.1,-0.05) {\footnotesize{$0$}};
  \node[black] at (1, .2) {\footnotesize $x$};
   \node[black] at (.45, 1) {\footnotesize $T_c <T<T_-$};
        
    \end{axis}
    \end{tikzpicture}
    \end{minipage}
  \hskip -1 cm
    \begin{minipage}[b]{0.48\textwidth}
    \centering
    \begin{tikzpicture}
    \begin{axis}[
        axis lines = middle,
        axis line style = {->, >=stealth, thick},
        ticks = none,
        xlabel = {${}$},
        ylabel = {${}$},
        xlabel style = {anchor=north west},
        ylabel style = {anchor=south east},
        xmin = -0.05, xmax = 1.05,
        ymin = -2.5, ymax = 3.5,
        width = 6.8cm,
        height = 5.5cm
    ]
        \addplot [
            domain=0:1.01,
            samples=200,
            smooth,
            color=blue!70!black,
            thick
        ] { (18 + 450*x^2 - 1150*x^85) * x * (x-0.7)^2 * (0.93-x) };

        \node[black, below left] at (axis cs:.1,-0.08) {\footnotesize{$0$}};
        \node[black] at (1.03, .4) {\footnotesize $x$};
        \node[black] at (.45, 2.6) {\footnotesize $T=T_c$};
    \end{axis}
    \end{tikzpicture}
    \end{minipage}
\caption{\small{Plots of the left hand side of \eqn{Txx} for $T_c<T<T_-$ (left) and precisely at $T=T_c$ (right), for $c_a>1$ and $c_b>1$.
As $T \gtrsim  T_-$ the nontrivial solution $x\gtrsim 0$.}}
\label{fig3}
\end{figure}

\no
If  one of the parameters $c_a$ and $c_b$ is smaller than 1 (but still positive) then the temperature $T_c$ does not exist and the situation is the same as that depicted on the left of fig.~\ref{fig3}, extending to all temperatures $0<T<T_-$.

\no
At temperatures below $T_c$, two additional solutions appear, as depicted in fig. \ref{fig4}.

\pgfplotsset{compat=newest}

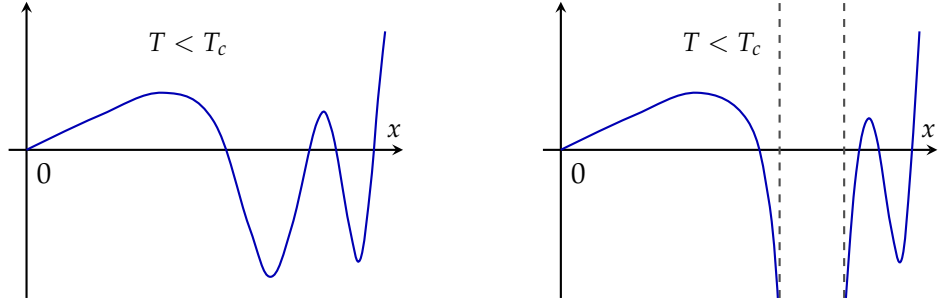
\begin{figure}[htbp]
\centering
    \begin{minipage}[b]{0.48\textwidth}
    \centering
    \begin{tikzpicture}
    \begin{axis}[
        axis lines = middle,
        axis line style = {->, >=stealth, thick},
        ticks = none,
        xmin = -0.05, xmax = 1.05,
        ymin = -3.5, ymax = 3.5,
        width = 6.8cm,
        height = 5.5cm
    ]
        \addplot [
            smooth,
            tension = 0.65,
            color = blue!70!black,
            thick
        ] coordinates {
            (0.00,  0.00)
            (0.20,  0.80)
            (0.38,  1.35) 
            (0.52,  0.70)
            (0.62, -1.80) 
            (0.68, -3.00) 
            (0.74, -1.80)
            (0.81,  0.65) 
            (0.85,  0.50)
            (0.90, -1.80) 
            (0.93, -2.60) 
            (0.96, -0.80)
            (0.98,  1.20) 
            (1.00,  2.80)
        };
        
        \node[black, below left] at (axis cs:.1,-0.05) {\small $0$};
        \node[black] at (1.025, .45) {\footnotesize $x$};
        \node[black] at (.45, 2.5) {\footnotesize $T<T_c$};

    \end{axis}
    \end{tikzpicture}
    \end{minipage}
    \hskip -1 cm
    \begin{minipage}[b]{0.48\textwidth}
    \centering
    \begin{tikzpicture}
    \begin{axis}[
        axis lines = middle,
        axis line style = {->, >=stealth, thick},
        ticks = none,
        xmin = -0.05, xmax = 1.05,
        ymin = -3.5, ymax = 3.5,
        width = 6.8cm,
        height = 5.5cm
    ]
        \draw [dashed, black!70, thick] (axis cs:0.61, -3.5) -- (axis cs:0.61, 3.5);
        \draw [dashed, black!70, thick] (axis cs:0.79, -3.5) -- (axis cs:0.79, 3.5);

        \addplot [
            smooth,
            tension = 0.65,
            color = blue!70!black,
            thick
        ] coordinates {
            (0.00,  0.00)
            (0.20,  0.80)
            (0.38,  1.35)
            (0.52,  0.70)
            (0.58, -1.20)
            (0.605,-3.50)
        };

        \addplot [
            smooth,
            tension = 0.65,
            color = blue!70!black,
            thick
        ] coordinates {
            (0.795,-3.50)
            (0.82, -0.80) 
            (0.85,  0.65) 
            (0.88,  0.30) 
            (0.92, -1.80)
            (0.95, -2.60) 
            (0.975,-0.50)
            (1.00,  2.80) 
        };
        \node[black, below left] at (axis cs:.1,-0.05) {\small $0$};
        \node[black] at (1.025, .45) {\footnotesize $x$};
        \node[black] at (.45, 2.5) {\footnotesize $T<T_c$};
   
    \end{axis}
    \end{tikzpicture}
    \end{minipage}

\caption{\small{Characteristic plots of the left hand side of \eqn{Txx} for $T<T_c$, for $c_a>1$ and $c_b>1$.}}
\label{fig4}
\end{figure}


\no
$\bullet$ \underline {$c_a c_b<1$:} Then $T_+>0$ and $T_-<0$.
The situation is the same as that depicted in fig. \ref{fig2} but with the
right figure valid for $0<T<T_+$.

\no
$\bullet$ \underline {$c_a,c_b<0,~ c_a c_b>1$:} Then both $T_\pm<0$. This is the purely paramagnetic case,
and it corresponds to the left fig.~\ref{fig2} but valid for all temperatures.

\no
Turning our attention to stability, the conditions \eqn{stabb4} assume the form (for $c=1$)
\be
\begin{split}
& \bigg({T\ov 1-x^2} -{c_a\ov 2} \bigg) \bigg({T\ov 1-y^2} -{c_b\ov 2} \bigg)>{1\ov 4} \ ,
\\
&
T >  { c_a\ov 2 } (1-x^2) \quad  \big({\rm equivalently}\ T >  { c_b\ov 2 } (1-y^2) \big)\ .
\end{split}
\label{stabeq}
\ee
From the above,  it becomes clear that the trivial (singlet) state, corresponding to $x=y=0$, is only stable for $T>T_+$, and it is the only state for such temperatures.
For $T<T_+$, the trivial state becomes unstable and there is a single stable solution with $x,y>0$. Below $T=T_-$, a new solution with $x>0$ appears (left figure in \eqn{fig3}),
but in fact it is unstable, so $T_-$ is not a critical temperature. Finally, for $T<T_c$, two additional solutions appear (fig. \eqn{fig4}). Of those, only the one for which
the curve has a positive slope is stable. The system has two locally stable solutions, the one with the larger value for $x$ being globally stable and the
other one metastable. For the globally stable solution the corresponding value for $y$ is also positive, implying that the magnets are polarized in the same direction, whereas for the metastable one $y$ is negative.

\no
We can understand the existence of these three phases by examining the system at zero temperature.
The free energy is simply the energy, found by putting $T=0$ and $j_a = j_b = \half$ (and $c=1$) in \eqn{F22free}
\be
\label{exy}
E(x,y)=-{1\over 4} (c_a x^2 + c_b y^2 + 2xy) \ ,
\ee
As argued, the range of variables can be taken to be
 $0\leqslant x\leqslant 1$ and $-1\leqslant y\leqslant 1$. The parameters $c_a$ and $c_b$
can have any sign, and, depending on their values, the signs of the parameters $T_\pm$ are fixed as depicted in fig.~\ref{fig1}. 

\no
Consider first the case $c_a c_b>1$. 
Then, when $c_a,c_b<0$, the point $x=y=0$ is a global minimum. If $c_a,c_b>0$, the point $x=y=0$ is 
a global maximum so that the minima of \eqn{exy} should occur either at the edges or at  the corners of the parametric space.
Since $c_a,c_b>0$, no minima may occur on the edges $(1,y)$ and $(x,\pm 1)$
and the global minimum occurs at $(1,1)$. The corner $(1,-1)$ is a local minimum, provided that 
 $c_a$ and $c_b$ are both larger than one. Altogether,
 \be
 \begin{split}
 c_a c_b >1: \quad &  {\rm global\ minimum\ at}\ (1,1)\ ,  {\rm if\ c_a>0\ \&\ c_b>0}\ .
 \\
 &  {\rm local\ minimum\ at}\ (1,-1) \  {\rm if\ c_a>1\ \&\ c_b>1}\ ,
 \\
 &   {\rm global\ minimum\ at}\ (0,0) \  {\rm if\ c_a<0\ \&\ c_b<0}\ .
 \end{split}
 \ee
 and their equivalents under $(x,y) \to (-x,-y)$. 
 Consider next the case $c_a c_b<1$. Then the point $x=y=0$ is a saddle point and there is only a single global minimum, either at the corners or on the
 edges of the parameter space. In particular, 
  \be
 \begin{split}
c_a c_b <1:\ \ {\rm global\ minimum\ at} \quad &  (1,1)\ {\rm if}\ c_a, c_b>-1\ ,
 \\
  \quad  & (1,-1/c_b) \ {\rm if}\ c_a>-1, c_b<-1\ ,
  \\
  \quad  & (-1/c_a,1) \ {\rm if}\ c_a<-1, c_b>-1\ ,
 \end{split}
 \ee
In all cases, $xy>0$ for the global minimum ($\neq(0,0)$) and $xy<0$ for the local one.

\begin{figure}[htbp]
    \centering
    \begin{tikzpicture}[>=latex, scale=.8]
        \fill[blue!20] (-3.5, 0) rectangle (0, 3.5); 
        \fill[blue!20] (0, -3.5) rectangle (3.5, 0); 
        \fill[blue!20] (0,0) -- (0,3.5) -- (1, 3.5) -- (1,1) -- (3.5, 1) -- (3.5,0) -- cycle;
        \fill[blue!20] (0,0) -- (-3.5,0) -- (-3.5,-0.285) -- plot[domain=-3.5:-0.285, samples=100] (\x, {1/\x}) -- (-0.285, -3.5) -- (0, -3.5) -- cycle;

        \fill[gray!20] (1, 1) rectangle (3.5, 3.5);
        
        \draw[thick, ->] (-3.5, 0) -- (3.5, 0) node[below] {$c_b$};
        \draw[thick, ->] (0, -3.5) -- (0, 3.5) node[left] {$c_a$};
        
        \node[below right] at (0,0) {\footnotesize  $0$};
        
        \draw[blue, thick, domain=0.28:3.2, samples=100] plot (\x, {1/\x});
        \draw[blue, thick, domain=-3.2:-0.28, samples=100] plot (\x, {1/\x});
        
        \draw[thick, dashed] (1, 1) -- (3.5, 1) node[above] {\small $c_a=1$};
        \draw[thick, dashed] (1, 1) -- (1, 3.5) node[right] {\small $c_b=1$};
        
        \node[blue, right] at (.19, .28) {\footnotesize $c_a c_b = 1$};
        \node[blue, left]  at (-1.5, -0.95) {\footnotesize $c_a c_b = 1$};
        
        \node[font=\small] at (2.2, 2.0) {double};
                 
        \node[font=\small] at (-2, -2) {trivial};
        
        \node[font=\small] at (-1.5, 1.5) {single};
    \end{tikzpicture}
    \vskip -.2 cm
    \caption{\small{Regions in the $(c_a\!-\!c_b)$-plane demarcating the double (grey), single (blue), and trivial (white) magnet areas.}}
    \label{fig11}
\end{figure}
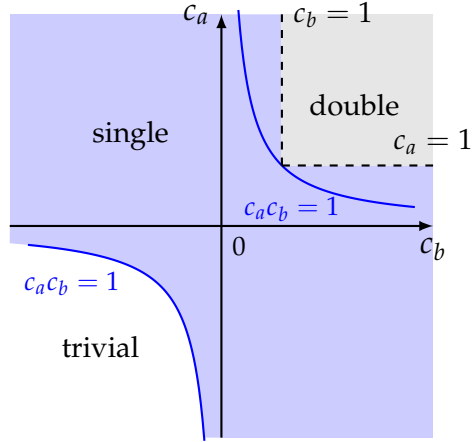

\no
In summary, if $c_a>1$ and $c_b>1$ the point $(1,1)$ is a global minimum and that in $(1,-1)$ is a local minimum. We call this the double magnet case.
 If $c_a c_b>1$ with both $c_a<0$ and $c_b<0$, the point (0,0) is the only minimum. This is the trivial (purely paramagnetic) case. 
 In all other cases the point $(1,1)$, or $(1,-1/c_b)$ or $(-1/c_a,1)$ is the only minimum. We call this the single magnet case. These
situations are depicted in fig.~\ref{fig11}, which is a refinement of fig.~\ref{fig1}.

\no
Increasing the temperature deforms the above configurations to the equilibrium states found previously. For instance, consider 
the global minimum state $(1,1)$ for $c_a c_b>1$ . For small nonzero $T$ this is deformed to
\be
{\rm deformed}\ (1,1): \quad x = 1-2 e^{-(c_a+1)/T} + \dots \ ,\quad y = 1-2 e^{-(c_b+1)/T} + \dots  \ .
\ee
Similarly, if  $c_a>1$ and $c_b>1$ the local minimum state for $(1,-1)$ is deformed to
\be
{\rm deformed}\ (1,-1): \quad x = 1-2 e^{-(c_a-1)/T} + \dots \ ,\quad y = -1+2 e^{-(c_b-1)/T} + \dots  \  .
\ee
Eventually, further increasing the temperature eliminates the deformed $x=-y=1$ state at $T=T_c$ for $c_a, c_b>1$,
and the deformed $x=y=1$ state at $T=T_+$.
The state $x=y=0$ is not deformed.
Note that the relative sign of $x$ and $y$ in nontrivial states is preserved, since neither variable can cross zero while the
other one remains nonzero.

\no
The limiting case $c_a c_b =1$ is interesting. 
In this case, the coupling matrix has a zero eigenvalue and, according to section \ref{decoup}, the system reduces to the mode corresponding to the eigenvector
of the nonzero eigenvalue and a subordinate mode for the eigenvector of the zero eigenvalue. Indeed, setting $c_b = 1/c_a$, \eqn{xyhalf} imply
the single equation of one variable
\be
z = c_a\tanh{z\over 2T} + \tanh{z\over 2c_a T}\ , \qq \text{with}\ \ z=c_a x+y \ ,
\label{z}
\ee
from the solution of which the individual values of $x$ and $y$ are found as
\be
x= \tanh{z \over 2T} \ ,\quad y = \tanh{z \over 2c_a T}\ .
\ee
Equation \eqn{z} has the solution $z=0$, as well as a nonzero solution $\pm z$ if $c_a>0$ and $T<\ha(c_a+1/c_a) = T_+$ (note that $T_-=0$,
so that there is no $T_c$), behaving for $c_a>0$ as a single
ferromagnet with an effective magnetization $z$ and a modified Brillouin function (cf. eq. \eqn{sssi} with $j=\ha$)
\be
B_{\ha,{\rm mod}} (s) =  c_a \tanh s + \tanh (s/c_a) \ .
\ee
The magnetizations $x$ and $y$ have the same sign and are fixed by the effective parameter $z$.

\no
It is interesting to follow the transition from the single magnet to the trivial one as we cross the curve $c_a c_b=1$ for negative $c_a$ (and $c_b$). 
We set $c_a c_b =1 -\e$, with $\e>0$ so we approach the the curve from the single magnet side (blue area in\ fig. \ref{fig11}) and take the limit $\e\ll 1$. 
The critical temperature $T_+$ is, to leading order in $\e$,
\be
T_+ = {\e |c_a |\over 2( c_a^2 + 1)}\ .
\equiv \e\, \tau_+
\ee
Therefore, the temperature range of the magnetized phase shrinks to $0<T<\e\tau_+$. Setting $T = \e \tau$, 
the second equation in \eqn{Txx} shows that, to leading order in $\e$, $y=-c_a x = |c_a| x$, and substituting in the first one we obtain
\be
\label{xbvcv}
x = \tau \bigg(\ln {1+|c_a| x\ov 1-|c_a| x}  + {1\ov |c_a|}  \ln {1+ x\ov 1-x} \bigg)\ .
\ee
This has a nontrivial solution $x\neq 0$ provided that $0<\tau < \tau_+$, as expected.
As $\tau \to \tau_+$, $(x,y)$ goes to zero, while as $\tau \to 0$, $(x,y)$ approaches $(1,|c_a|)$ (for $|c_a|<1$) or $(|c_a|^{-1},1)$ (for $|c_a|>1$).
For $c_a c_b >1$ (on the trivial side of the curve) the only state is $x=y=0$.

\subsubsection{The magnetization axis }
The signs of $x$ and $y$ are irrelevant for the spins of their respective irreps. As stated below \eqn{xyb}, the spins of the individual magnet states are
\be
J_a = j_a |x| n = \ha  |x| n \ , \quad
J_b = j_b |x| n = \ha  |y| n \ , 
\ee
where in the last step we restricted the spins of the atoms to $j_a=j_b=\ha$.
However, their relative sign affects the irrep of the total spin and has physical significance. The thermodynamically dominant irrep of the total spin is the one
with $\kk = \kk_a + \kk_b$ (see appendix \ref{partwo}), that is,
\be
k_i =k_{a,1}+k_{b,i}\ ,\qq i=1,2\ . 
\ee
Therefore, the total spin is
\be
J ={ |k_1-k_2|\ov 2} = \ha |x+y| n\ .
\ee
If $x$ and $y$ have the same sign, then $x+y = |x|+|y|$ and $J=J_a + J_b$. In this case, there is an axis along which $J_a^z = J_a$, $J_b^z = J_b$, 
$J^z = J_a^z + J_b^z = J_a + J_b$. This is the magnetization axis, and spins $J_a$ and $J_b$ are parallel along this direction. 

\no
If, however, $x$ and $y$ have opposite signs, then $|x+y|=|x| - |y|$ (assuming $|x| > |y|$, thus $J_a > J_b$), giving $J = J_a - J_b$.
In this case, $J_a^z = J_a$, $J_b^z = -J_b$, $J^z = J_a - J_b$ along the magnetization axis, and spins $J_a$ and $J_b$ are antiparallel.
This is a key physical difference between globally stable states, which have the maximum value of total angular momentum,
and metastable states, which have the minimum angular momentum. The corresponding phases of the magnet
are summarized in table \ref{taaa}.

\no
Note that a state with $x=-y\neq 0$ corresponds to $J_a = J_b = n|x|/2$ but has total spin $J=0$, and is a singlet with respect to global $SU(2)$ rotations.
Nevertheless,
the total spins of atoms $a$ and $b$ are nonzero, so the atoms are magnetized, with only the total magnetization being zero. To distinguish this state from
the state $x=y=0$ where atoms $a$ and $b$ are unmagnetized, we called this last state the trivial state. Experimental probes with differential sensibility to the 
magnetizations of atoms $a$ and $b$ would be able to distinguish between the trivial and the nontrivial singlet states.

\begin{table}[h]
\centering
\renewcommand{\arraystretch}{1.5}
\begin{tabular}{|c||c|c|c|}
\hline
Irrep$\backslash$Temperature & \textbf{$T < T_c$} & \textbf{$T_c < T < T_+$} & \textbf{$T > T_+$} \\ \hline\hline
$\bullet$ &  unstable  & unstable & stable \\ \hline
$\mathrel{\substack{\pmb{\longrightarrow} \\[-0.55ex] \hskip -.18 cm \pmb{\rightarrow}}} $& stable & stable & $\times$ \\ \hline
$\mathrel{\substack{\pmb{\longrightarrow} \\[-0.8ex] \hskip -.75 cm \pmb{\leftarrow}}}$  & metastable & $\times$ & $\times$ \\ \hline
\end{tabular}
\vskip -0.1cm
\caption{\small{Phase structure (locally stable configurations)
 when both $c_a>1$ and $c_b>1$ (gray region in fig.~\ref{fig11}).
When one of  $c_a, c_b$ is less than one but  still positive, 
or when $c_a c_b <1$ (blue region in fig.~\ref{fig11}),
the column with $T<T_c$ as well as the last row should be removed and the next column extends to $0< T < T_+$.
When both $c_a$ and $c_b$ are negative and $c_a c_b >1$ (white region in fig.~\ref{fig11}),
the trivial (singlet) is the only stable solution for all temperatures. The arrows denote the (generally unequal) magnetizations of the atoms,
either along the same or in opposite directions. 
}}
\label{taaa}
\end{table}
\no
Compared to the single magnet case ($m=1$), an additional metastable phase appears below the critical temperature $T_c$ for the double magnet,
corresponding to the two magnets polarized in opposite directions. It is expected that adding magnets will increase the number of metastable phases
below some critical temperatures.

\subsubsection{Phase transitions}
The system undergoes a second order phase transition at $T=T_+$. To see this, we examine equation \eqn{Txx} at temperatures
near $T_+$. For $T>T_+$ the state is the trivial one $x=y=0$, while for $T<T_+$ it is the magnetized state $(x,y)$. For $T \lesssim  T_+$ it is given by
\be
\label{fihi2}
x \simeq x_+ (T_+-T)^{1/2}\ ,\qq x_+  = \sqrt{ {3( T_+-T_-)}\ov T_+(T_+-c_a/2) (\D-2)}\ .
\ee
where $\D = (c_a-c_b)^2+4$ as in \eqn{tpmd}.
The corresponding expression for $y$ is as above, but with $c_a$ replaced by $c_b$. 
The free energy for the $j_1=j_2=\ha $ case has the form
\be
F =  - 2 T \ln 2 + 
{c_a\ov 4} x^2 +  {1 \ov 2}x y + {c_b\ov 4}  y^2
- T \ln \left( \cosh\hb\frac{c_a x + y}{2T}\,  \cosh\hb\frac{c_b y + x}{2T}  \right)\ 
\ee
and using \eqn{fihi2} we obtain
\be
\begin{split}
 & F = - 2 T \ln 2   - {3 \D\ov 4 (\D-2) T_+ } (T_+-T)^2 + \dots \ ,\qq {\rm for}\ T\lesssim T_+\  ,
 \\
 &
  F =  - 2 T \ln 2  \ ,\qq {\rm for}\ T > T_+\ .
 \end{split}
\ee
We see that there is a discontinuity in the second derivative of the free energy at $T=T_+$, marking a second-order phase transition.

\no
The transition at $T=T_c$ is different, as it represents a delicate hysteresis phenomenon. As the temperature drops
from $T_+$ to zero, the state remains the one with parallel magnetization. At a temperature $T<T_c$, the system
can be prepared to be in the state with antiparallel magnetization. This can be achieved by applying external fields that couple differently to the spins
of atoms $a$ and $b$ such that the antiparallel state be selected, and slowly driving the fields to zero. The antiparallel state is
metastable for $T<T_c$ and, unless externally perturbed, the magnet will remain in this state as the temperature is raised. At $T=T_c$ the antiparallel
state destabilizes and the magnet will transit to the parallel state, with a lower free energy. This transition, happening only when the temperature
is raised, can be called zeroth-order, as there is a discontinuity in the free energy.

\subsubsection{An instructive special case: equal couplings}

In the case $c_a = c_b \equiv c_0$ the equilibrium equations and stability conditions simplify and we can obtain analytical results.
The critical temperatures $T_\pm$ become
\be
T_\pm = {c_0 \pm 1 \over 2} \ .
\ee
The equilibrium equations \eqn{xyhalf} become symmetric under $x \leftrightarrow y$ and under $x \leftrightarrow -y$, and we can look for
solutions invariant under these transformations, namely $x=y\equiv x_+$ and $x=-y \equiv x_-$ (the latter is a nontrivial singlet state). They obey the equation
\be
T \ln{1+x_\pm \over 1-x_\pm} = (c_0 \pm 1) x_\pm = 2 T_\pm x_\pm \ .
\label{eqst}\ee
These are identical to the equations for a single spin-$\half$ magnet with Curie temperatures $T_\pm$. Magnetized states $x_\pm \neq 0$
will exist for $T<T_+$ for $x_+$, and for $T<T_-$ for $x_-$, and correspond to the parallel ($x_+$) and antiparallel ($x_-$) magnetization phases.
The stability condition \eqn{stabeq} in this case becomes
\be
T > (1-x_\pm^2) {c_0 +1\over 2} = (1-x_\pm^2) T_+
\label{Tstpm}\ee
Note that only $T_+$ appears, since it demarks the largest eigenvalue of the stability matrix.

\no
Each phase will be stable if the corresponding condition \eqn{Tstpm} is satisfied. Expressing the temperature in terms of $x_\pm$ by using \eqn{eqst}, we can
state the stability condition as
\be
{1-x_\pm^2 \over 2x_\pm} \ln{1+x_\pm \over 1-x_\pm} < {T_\pm \over T_+}
\ee
The function on the left side of the condition varies from 1 to 0. Therefore, for $x_+$ it is identically satisfied, and the state $x=y$ is stable for all
temperatures $T<T_+$. For $x_-$, however, $T_-/T_+ <1$ and $x_-$ must be above a critical value $x_c$ for stability to hold. The critical temperature
$T_c$ below which the state $x=-y$ stabilizes is given by the solution of the transcendental equation
\be
{1-x_c^2 \over 2x_c} \ln{1+x_c \over 1-x_c} = {c_0-1\over c_0+1} \ , \qq T_c = {(c_0-1)x_c \over \ln{1+x_c \over 1-x_c}}={2x_c \over \ln{1+x_c \over 1-x_c}} T_-
\ee
Note that $T_c <T_-$.

\no
So, for $c_0 >1$ the phase transition scenario is as in table \ref{taaa} 
(the part $T_c<T<T_-$ with the physically irrelevant unstable antiparallel magnetization state is not shown in the table). For $-1<c_0<1$, $T_-$ and $T_c$
disappear and we only have the trivial state for $T>T_+$ and the parallel magnetization state below $T_+$. And for $c_0<-1$, $T_+$ also disappears and
we only have the trivial state.

\begin{figure}[htbp]
\centering
\begin{tikzpicture}
\begin{axis}[
    axis lines = middle,
    axis line style = {->, >=stealth, thick},
    ticks = none,
    xticklabels = \empty,
    yticklabels = \empty,
    xlabel = {\footnotesize{$x$}},
    xlabel style = {anchor=north west},
    ylabel style = {anchor=south east},
    xmin = -0.1, xmax = 2.15,
    ymin = -0.12, ymax = 0.33,
    width = 6.0cm,
    height = 5.5cm,
    clip = false
]
 \node[black, below left] at (axis cs:1.5,0.3) {\footnotesize{$T=T_c$}};
    \addplot [
        smooth,
        tension = 0.65,
        color = blue,
        thick
    ] coordinates {
        (0.00, 0.000)
        (0.15, 0.190)
        (0.28, 0.230)
        (0.50, 0.100)
        (0.75, 0.020)
        (0.95, 0.002)
        (1.05, 0.000)   
        (1.20, -0.005)  
        (1.35, -0.020)
        (1.50, -0.055)
        (1.68, -0.090)  
        (1.82, -0.045)
        (1.90, 0.000)   
        (2.02, 0.140)
        (2.10, 0.310)
    };

\end{axis}
\end{tikzpicture}%
\hspace{0.9cm}%
\begin{tikzpicture}
\begin{axis}[
    axis lines = middle,
    axis line style = {->, >=stealth, thick},
    ticks = none,
    xticklabels = \empty,
    yticklabels = \empty,
    xlabel = {\footnotesize{$x$}},
    xlabel style = {anchor=north west},
    ylabel style = {anchor=south east},
    xmin = -0.1, xmax = 2.15,
    ymin = -0.08, ymax = 0.32,
    width = 6.0cm,
    height = 5.5cm,
    clip = false
]
 \node[black, below left] at (axis cs:1.5,0.284) {\footnotesize{$T\lesssim T_c$}};
    \addplot [
        smooth,
        tension = 0.65,
        color = blue,
        thick
    ] coordinates {
        (0.00, 0.000)
        (0.10, 0.150)
        (0.25, 0.213)
        (0.45, 0.110)
        (0.65, 0.030)
        (0.88, -0.005)
        (1.15, 0.005)
        (1.35, -0.015)
        (1.67, -0.070)
        (1.90, 0.000)
        (2.02, 0.120)
        (2.10, 0.285)
    };

\end{axis}
\end{tikzpicture}
\caption{\small{Plot of the left hand side of \eqn{Txx} in the case $c_a=c_b$ for $T=T_c$ (left) and $T\lesssim T_c$ (right). 
For $T\lesssim T_c$ the slope at $x=x_c$ becomes positive, giving a stable solution flanked by two unstable ones,
all stemming from the same unstable solution at $T=T_c$. This is unlike the case in fig. \ref{fig4}, where for 
$T\lesssim T_c$ the emerging stable and unstable solutions stem from a point different from the existing unstable solution.}}
\label{figab}
\end{figure}
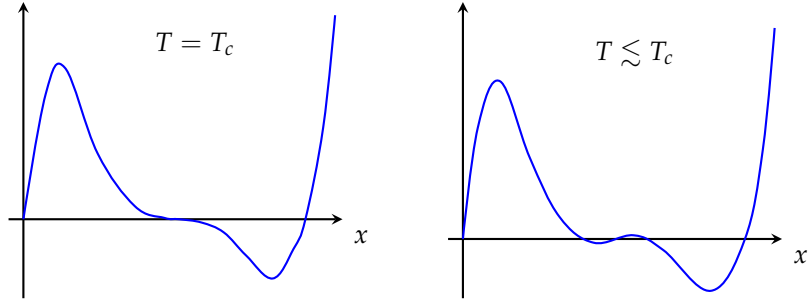

\no
The general analysis of section \ref{su2h} remains valid in the case $c_a=c_b$, with the difference that the stable antiparallel configuration
does not appear as a new state at $T=T_c$, as in the right panel of fig.~\ref{fig3}, but rather as a continuation of the unstable one. 
We can show that,
in this case, if the function in the left hand side of \eqn{Txx} and its derivative vanish (a condition for being at $T=T_c$), the second derivative also vanishes
and the picture changes as in fig.~\ref{figab}.
\vskip -0.1cm
\subsection{The  $SU(2)$ case for general spin}

For general $j_a,j_b$, we define $T_\pm$ as
\be
T_\pm = {1\ov 3} \Big(
c_a j_a (1 + j_a) + c_b j_b (1 + j_b)
\pm  \sqrt{\Delta_{j_1,j_2}} \Big)\ ,
\ee
where 
\be
\Delta_{j_a,j_b} = \big( c_a j_a (1 + j_a) - c_b j_b (1 + j_b) \big)^2
+ 4 c^2 j_a j_b (1 + j_a) (1 + j_b)\ . 
\ee
It turns out that the results for the case $j_a=j_b=\ha$ remain qualitatively unchanged for the general spin case. This includes the case in which 
one or both of the spins become infinite. 
We believe that the reason for that is the fact that the Brillouin function $B_j$ does not change in general shape and properties as the spin 
ranges from $\ha$ to infinity.
The expressions for the expansion of the free energy near $T=T_+$ are quite complicated for the
general case $j_a\neq j_b$ and will not be presented here. We only give the result for the equal spin case $j_a=j_b=j$ (and $c=1$),
\be
\begin{split}
  & F = - 2 T \ln (2 j+1)   - {5j(j+1)\D\ov 2 (1+2 j+ 2 j^2)(\D-2)T_+} (T_+-T)^2 + \dots \, ,\quad {\rm for}\ T\lesssim T_+\  ,
 \\
 &
  F =  - 2 T \ln (2 j+1)  \, ,\quad {\rm for}\ T > T_+\ .
 \end{split}
\ee
where  $\Delta = 4 + (c_a - c_b)^2$ as in \eqn{tpmd}, revealing a second order phase transition.

\no
In the case that both $j_a,j_b\to \infty$ one has to appropriately scale $c_a\to c_a/j_a^2$, $c_b\to c_b/j_b^2$, and $c\to c/(j_a j_b)$
to have a nontrivial large spin limit. Then the free energy is 
\be
F =  c_a x^2 + 2 c x y + c_b y^2 
- T \ln\left( \frac{ T^2 \sinh \frac{2 (c_a x + c y)}{T} \sinh {2 (c x + c_b y)\ov T}}{(c_a x + c y)(c x + c_b y)}
\right)\ ,
\ee
where we have discarded a constant proportional to the logarithm of the spins, 
and for the singlet 
\be
F_{\rm singlet} =  - 2 T \ln 2 \ .
\ee 
Then (for $c=1$)
\be
\begin{split}
 & F = - 2 T \ln 2   -
\frac{15 \Delta}
{4 (\Delta - 2) (c_a+c_b+\sqrt{\D})}(T_+-T)^2 + \dots \ ,\qq {\rm for}\ T\lesssim T_+\  ,
 \\
 &
  F =  - 2 T \ln 2  \ ,\qq {\rm for}\ T > T_+\ , 
 \end{split}
\ee
where $T_\pm =(c_a+c_b\pm \sqrt{ \D})/3$, again revealing  a second order phase transition. 

\no
Note also the interesting case with only mutual interaction, i.e.  $c_a=c_b=0$. Then, the equilibrium equations \eqn{xyb} become
\be
x= B_{j_1}(2 j_aj_b c y/T)\ ,\qq y= B_{j_b}(2 j_aj_b c x/T)\ .
\ee
Hence in this case we may obtain the full solution from the single equation
\be
x   =B_{j_a}\bigg({2 j_aj_b c\ov T} B_{j_b}\Big({2 j_aj_b c x\ov T}\Big)\bigg)\ .
\ee
The phase transitions remain second order, as the limit $c_a,c_b\to 0 $ in the corresponding formulae remains finite.

\section{Concluding remarks}\label{conclusions}

\no
The inclusion of many magnetic components within an elementary magnet ("atoms" within "molecules") enriches the thermodynamical structure of the resulting
macroscopic magnet with new phases and corresponding transitions. The most striking effect is the emergence of fully ferromagnetic,
paramagnetic, and mixed ferro-antiferromagnetic behavior depending on the structure and eigenvalues of the coupling matrix between components.

\no
In fact, we have just scratched the surface of this structure in the present paper. Due to the nontrivial, coupled nature of the equilibrium equations and
the impossibility to derive exact stable solutions, our numerical analysis was confined to the simplest nontrivial case of two spin-$j$ components of $SU(2)$ magnets,
providing a generalization of the usual ferromagnet. Even in this minimal case, nontrivial properties emerged and the behavior of the magnet can be qualitatively
different from the standard case, with either a singe or two distinct magnetization parameters, up to two critical temperatures, and corresponding metastability
properties.

\no
There are many obvious directions for future investigation. The possibilities are manifold (in fact, almost forbidding), involving more than two magnetic components in higher representations of higher $SU(N)$ groups. Further progress and research efforts have to be guided by the physical interest of the model and its proximity
to realistic magnetic systems. One such model, involving many identical spin-$j$ $SU(2)$ components with a sort of nearest-neighbor interactions, is
studied in a companion paper \cite{toallo}. This can be relevant to magnetically active macromolecules. Other possibilities exist and will be the subject of
future research.


\subsection*{Acknowledgements }

 A.P.'s research was supported in part by the National Science Foundation 
under grant NFS-PHY-2112479, and by PSC-CUNY grants 69213-00 57 and 6D136-00 06.

\appendix
\section{Microscopic Hamiltonian for two atoms}
\label{microhami}

 \no
To illustrate the concept of interacting atoms and a global $SU(N)$ symmetry with a specific example, consider molecules consisting of two atoms $A,B$ with $N$ states each, for a total of $N^2$ internal states.
 Denoting the states of molecule $i$ as $\ket{a, b}_i$, $a\bb \in\bb A$, $b\bb\in\bb B$, $a,b = 1,\dots,N$, we take the interaction Hamiltonian to be (direct product symbols
 between states of different molecules are omitted)
 \ba
 H = && \hskip -0.5cm \half \sum_{i,j,a,a',b,b'} \bb\Big( c_{A,ij} \ket{a' ,b}_i \ket{a ,b'}_j  \bra{a ,b}_i  \bra{a' ,b'}_j  + c_{B,ij} \ket{a,b'}_i \ket{a',b}_j \bra{a,b}_i \bra{a' ,b'}_j \cr
  &&{}\hskip 1cm +
 c_{AB,ij} \ket{b' ,b}_i \ket{a' ,a}_j  \bra{a ,b}_i  \bra{a' ,b'}_j \\ 
  &&{}\hskip 1cm+ {\tilde c}_{A,ij} \ket{a' ,b}_i \ket{a' ,b'}_j  \bra{a ,b}_i  \bra{a ,b'}_j  
  + {\tilde c}_{B,ij} \ket{a,b'}_i \ket{a',b'}_j \bra{a,b}_i \bra{a' ,b}_j \cr
 &&{}\hskip 1cm +  
   {\tilde c}_{AB,ij} \ket{b',b}_i \ket{a,b'}_j \bra{a',b}_i \bra{a,a'}_j  \Big)\ , 
 \nonumber\ea
 where $c_{A,ij} \bb=\bb c_{A,ji}$, $c_{B,ij}\bb=\bb c_{B,ji}$, $c_{AB,ij}$, $c_{BA,ij}$ and correspondingly for $\tilde c$ are real constants.
 
 \no
 The first three terms in $H$ are exchange interactions.
 The first two exchange the states $A$ of two molecules (first term, in which $a$ and $a'$ are interchanged) and the states $B$ of two molecules (second term, in which $b$ and $b'$ are interchanged) and have a global
 $SU(N)_A \times SU(N)_B$ invariance under which states $A$ and $B$ transform in the fundamental ($F$) representation of each group factor.
 The third term exchanges the sates $A$ and $B$ ($a$ and $b'$ are interchanged) of two molecules and has a global $SU(N)$ invariance under which states $A$ and $B$ transform in the
 representation $F \times F$.

 \no
 The last three terms represent projections on maximally entangled states. The fourth and fifth terms are projections on entangled states $A$
(fourth term) and entangled states $B$ (fifth term) between two molecules and have a global $SU(N)_B$ invariance (fourth term) or $\times SU(N)_A$
(fifth term) under which the corresponding states transform in the fundamental. Finally, the last
term is a projection on maximally entangled states $A$ of one molecule with states $B$ of another molecule and has a global $SU(N)$ invariance under
which states $A$ and $B$ transform in the representation $F \times {\bar F}$. More compactly, the Hamiltonian can be written as
\be
H = \half\sum_{ij} \big( c_{A,ij} P_{A,ij} + c_{B,ij} P_{B,ij} + c_{AB,ij} P_{AB,ij} +  {\tilde c}_{A,ij} \Pi_{A,ij} +  {\tilde c}_{B,ij} \Pi_{B,ij} +  {\tilde c}_{AB,ij} \Pi_{AB,ij}\big)\ ,
\ee
with $P$ and $\Pi$ corresponding permutation and projection operators respectively.

\no
The fully general Hamiltonian has no symmetries, as its parts have incompatible symmetries. Situations with a global $SU(N)$ symmetry
arise when some of the coefficients vanish. Special cases of interest are:

\no
$\bullet$ All coefficients $\tilde c$ vanish. Then the magnet has a global $SU(N)$ symmetry with states transforming in the representation $F \times F$. In terms of
fundamental $SU(N)$ generators $j_{A,i}^a$, $a=1,\dots,N^2-1$, acting on states $A$ of molecule $i$, and similarly for $j_{B,i}^a$, the Hamiltonian
can be written
\be
H = \sum_{i,j,a} \big( c_{A,ij} \,j_{A,i}^a\, j_{A,j}^a + c_{B,ij}\,  j_{B,i}^a\,  j_{B,j}^a + c_{AB,ij}\,  j_{A,i}^a\, j_{B,j}^a \big)\ .
\label{Hff}\ee

\no
$\bullet$  $c_{AB,ij} = {\tilde c}_{A,ij} = {\tilde c}_{B,ij} = 0$. Then the magnet has a global $SU(N)$ symmetry with states transforming in the representation
$F \times {\bar F}$. In terms of fundamental and antifundamental generators ${\bar j}_{A,i}^a = - j_{A,i}^{a*}$ and similarly for ${\bar j}_{B,i}^a$ the Hamiltonian can be written
\be
H = \sum_{i,j,a} \big( c_{A,ij} \,j_{A,i}^a\, j_{A,j}^a + c_{B,ij}\,  {\bar j}_{B,i}^a\,  {\bar j}_{B,j}^a + {\tilde c}_{AB,ij}\,  j_{A,i}^a\, {\bar j}_{B,j}^a \big) \  .
\label{Hfa}
\ee
The above Hamiltonians \eqn{Hff} and \eqn{Hfa} are special $m=2$ instances of the general Hamiltonian \eqn{Hgeneral} presented in section {\bf \ref{tbH}}.
 
 \section{Derivation of partition function for $m=2$}
 \label{partwo}
 
 Assume that each molecule has two atoms, and thus supports a representation of $SU(N)$ $A \otimes B$. Its Hilbert space is
 ${\cal{H}}_A \otimes {\cal{H}}_B$, the factors having dimensionality $\dim(A)$ and $\dim(B)$, on which generators $j_A^a$ and $j_B^a$ act.
 The most general interaction between molecules that respects $SU(N)$ invariance is, by using \eqn{Hgeneral}, given by
\be
H_{ij} = \ha c_{A�,ij} \sum_a j_{A,i}^a j_{A,j}^a +\ha c_{BB,ij} \sum_a j_{B,i}^a j_{B,j}^a + c_{AB,ij} \sum_a j_{A,i}^a j_{B,j}^a\ ,
\label{H2}\ee
with $c_{AA,ij}, c_{BB,ij}$ symmetric in $i,j$. The mean-field Hamiltonian in the $n\gg1$ limit is 
then given by \eqn{Hcoupled} with $m=2$ which we rewrite it as
\be
H =-{c_1\over 2 n} \sum_a \big(J_A^a\big)^2   - {c_2 \over 2 n} \sum_a  \big(J_{B}^a\big)^2
 - {c_3 \over 2 n} \sum_a\big(J_{A}^a + J_{B}^a\big)^2\ ,
\label{Hmf}
\ee
where the entries of the symmetric matrix $C_{st}$ are
\be
C_{AA}=c_1+c_3\ ,\qq C_{BB} = c_2 + c_3\ ,\qq C_{AB} = C_{BA} = c_3
\label{Cc}\ee
and $J_{A}^a$, $J_{B}^a$ are the global $SU(N)$ generators
\be
J_A^a = \sum_i j_{A,i}^a \ ,\qq  J_B^a = \sum_i j_{B,i}^a\ .
\label{H2}
\ee
Note that the first two terms in \eqn{Hmf} are invariant under $U(N)\times U(N)$, but the third
term reduces the symmetry of the full Hamiltonian to $U(N)$.

\no
The full Hilbert space consists of $n$ copies of irrep $A$ and $n$ copies of irrep $B$. We decompose it into irreps $\rr$ of $J_A^a$ and irreps $\ss$ of $J_B^a$,
and then decompose the irrep $\rr \times \ss$ into irreps $\kk$ of $J_A^a + J_B^a$. Calling $| d_{A,\rr} |$ the number of irreps $\rr$ in the tensor product
of $n$ irreps $A$,  $| d_{B,\ss} |$ the number of irreps $\ss$ in the tensor product
of $n$ irreps $B$, and  $| d_{\rr,\ss;\kk} |$ the number of irreps $\kk$ in the tensor product $\rr \times \ss$ we have
\be
\label{hhhp}
Z = \sum_{\rr,\ss,\kk} | d_{A,\rr} | \, | d_{B,\ss} | \, | d_{\rr,\ss;\kk} |\, \dim(\kk) \exp\Big({{\b c_1\over 2n} \rr^2 + {\b c_2\over 2n} \ss^2 +{\b c_3\over 2n}\kk^2}\Big)
\ee
in the obvious notation $\displaystyle \kk^2 = \sum_i k_i^2$ and similarly for $\rr$ and $\ss$. 
The absolute values in $| d_{A,\rr} |$ etc$.$ are put since we will define
$d_{A,\rr}$ antisymmetric in $\rr$, and similarly for $d_{B,\ss}$ and $d_{\rr,\ss;\kk}$. Their expressions are given by \eqn{dnrk} as 
\be
d_{A,\rr} = \oint {d\zz_A \over \zz_A^{\rr+1}} \,\D(\zz_A)\, \chi_A(\zz_{A})^n\ ,\qq d_{B,\ss} = \oint {d\zz_B \over \zz_B^{\ss+1}} \,\D(\zz_B)\, \chi_B(\zz_B)^n\ ,
\label{drds}
\ee
in the obvious notation $\displaystyle \zz_A^{\rr+1} = \prod_{i=1}^N z_{A,i}^{r_i+1}$ and 
$ \displaystyle d\zz_A = {1\ov (2\pi i)^N}  \prod_{i=1}^N dz_{A,i}$  and similarly for $d\zz_B$,
 while, using \eqn{psidchi}
\be
d_{\rr,\ss;\kk} = \oint {d\zz \over \zz ^{\kk+1}}    {\psi_r(\zz) \psi_s(\zz) \over \D(\zz)}\ ,
\label{rsk}
\ee
where $\psi_r(\zz)$ is the antisymmetrized Slater determinant \eqn{slater}
Finally, the dimensionality of the irrep $\bf k$
\be
\dim(\kk) = {|\D(\kk)| \over \prod_{s=1}^{N-1} s!} \ ,
\ee
appears in \eqn{hhhp}.
We now observe that, when used in sums over $\rr,\ss$ or $\kk$, the following identities hold 
\be
|d_{A,\rr}| \psi_r(\zz) = N!\, d_{A,\rr}\, \zz^\rr \ ,\quad |d_{B,\ss}|  \psi_r(\zz) = N!\, d_{B,\ss}\, \zz^\ss \ ,\quad
| d_{\rr,\ss;\kk} | \, |\D(\kk) | = d_{\rr,\ss;\kk} \, \D(\kk) \ .
\ee
Consequently,  putting everything together, we obtain (up to an irrelevant overall factor)
\be
\label{partii}
\begin{split}
 Z = \sum_{\rr,\ss,\kk} \oint & d\zz\, d\zz_A\, d\zz_B   {\D(\zz_A) \D(\zz_B) \over \D(\zz)}
{\zz^{\rr+\ss-\kk-1} \over
\zz_A^{\rr+1} \,\zz_B^{\ss+1}}\, \chi_{A}(\zz_a)^n \chi_B (\zz_b)^n\D(\kk) 
\\
& \times \exp\Big({{\b c_1\over 2n} \rr^2 + {\b c_2\over 2n} \ss^2 +{\b c_3\over 2n} \kk^2}\Big)\ ,
\end{split}
\ee
which is an exact expression. In the large-$n$ limit, the typical $\kk, \ss,\rr$ are of order $n$, so the exponent is of order $n$. In this limit,
the Vandermonde determinants of the various $\zz$ are of order $z^{N(N-1)/2}$ and are subleading compared to the powers of $\zz$, which are
of order $z^n$. Similarly, the Vandermonde $\D(\kk)$ is subleading compared to the exponential. Omitting them, the integration over $\zz$
gives $\delta(\rr+\ss-\kk)$. Putting $\rr = n {\bf x}_A$,  $\ss = n {\bf x}_B$, $\zz_{A,B} = e^{\bf w_{A,B}}$ and 
$\l_{A,B} ({\bf w}_{A,B}) = \ln \chi_{A,B} (e^{\bf w_{A,B}})$, we have 
\be
Z = \int d{\bf x}_A \, d\zz_A\, d{\bf x}_B \, d\zz_B  \, e^{-n \beta F}\ ,
\ee
with
\be
F = T{\bf x}_A {\bf w}_A + T{\bf x}_B {\bf w}_B - { c_1\over 2} {\bf x}_A^2 - {c_2\over 2} {\bf x}_B^2 -{c_3\over 2} ({\bf x}_A + {\bf x}_B)^2
-T \l_A ({\bf w}_A) -T \l_B ({\bf w}_B) \ .
\label{freeF}\ee
This is the $m=2$ version of \eqn{Fn} upon the identification \eqn{Cc} of coupling constants.

\no
Note that the relation $\kk = \rr + \ss$ would seem to imply that the composition of irreps $\rr$ and $\ss$ leads to a single irrep
$\rr + \ss$, which is obviously untrue. This is an artifact of the fact that we used the expression for $d_{\rr,\ss;\kk}$
inside the partition function sum, and that we made a large-$n$ approximation, omitting subleading terms.
This justifies the assertion below \eqn{jjjk}  that irreps with $\kk_s + \kk_t$ dominate in the thermodynamic free energy.
The expression \eqn{rsk} remains correct and exact (and also \eqn{partii}).

\section{Stability analysis}

Denoting collectively the $m N$ variables $x_{s,i}$ as the vector $\bX$, stability is determined by the positivity of the ($m N$)-dimensional matrix
${\tilde {\bf H}}$
\be
\label{c1}
{\tilde {\bf H}} = {\bf P} {\bf H} {\bf P}\ ,
\ee
where $\bf H$ is the Hessian of the system without constraints and $\bf P$ the projector on the space orthogonal to the constraints.
The constraints \eqn{cons} can be written as
\be
{\bf u}_s \cdot \delta \bX = n_s\ ,\qq ({\bf u}_s)_{s',i} = \delta_{ss'}\ ,\qq  {\bf u}_s \cdot {\bf u}_t = N \delta_{s,t}\ .
\ee
The constraint vectors ${\bf u}_s$ are orthogonal and the projector $\bf P$ assumes the form
\be
{\bf P}_{si,tj} = \delta_{st} \Big( \delta_{ij} - {1\over N}\Big)\ ,\qq  {\bf P} \,{\bf u}_s = 0
\ee
A direct calculation of the Hessian of the free energy \eqn{Ft} yields
\be
\label{c4}
{\bf H} = {\bf W} - \b\, {\bf C}\ ,
\ee
with
\be
{\bf W}_{si,tj} = W_{s,ij}\, \delta_{st}\ , \qq W_{s,ij} = {\partial w_{s,i}({\bf x}_s) \over \partial x_{s,j}}\ ,\qq
{\bf C}_{si,tj} = C_{st}\, \delta_{ij}\ .
\ee
Defining also the matrix
\be
{\bf \Lambda}_{si;tj} = \Lambda_{s,ij}\, \delta_{st}\ ,\qq \Lambda_{s,ij} = {\partial^2 \lambda_s ({\bf w}_s) \over \partial_{w_{s,i}} \partial_{w_{s,j}}}\ ,
\ee
the matrices ${\bf W}, {\bf \Lambda}, {\bf P}$ satisfy
\be
{\bf \Lambda} {\bf u}_s = 0\ ,\qq {\bf \Lambda}{\bf P} = {\bf P} {\bf \Lambda} = {\bf \Lambda} \ ,\qq  
 {\bf \Lambda}   {\bf W}  {\bf P}= {\bf P}\ 
\ee
and obviously ${\bf C}$ and $ {\bf P}$ commute. 
Then. \eqn{c1} gives $\bf \tilde H = {\bf P W P} - \bf P C P = ({\bf \L}^{-1}-\bf {\bf C}){\bf P}$.
Thus, on the projected space, ${\bf W}$ can be replaced by ${\bf \Lambda}^{-1}$ (with the zero modes ${\bf u}_s$ removed), so
\be
{\tilde {\bf H}}_{si;tj} = ({\Lambda}_s )_{ij}^{-1} \delta_{st} - \b\, {C}_{st}\, \delta_{ij}
~~~{\text{on states $\bf v$ such that}} ~~~{\bf P}\, {\bf v} = {\bf v}\ ,
\ee
recovering \eqn{hese}.
 
 \section{Equilibrium and off-equilibrium free energies}
 \label{relonoff} 

In this appendix we examine similarities and differences of the two free energies \eqn{Fe} and \eqn{Ft} in a more general context.

\no
The free energy \eqn{Fe} is in principle valid only for equilibrium configurations $x_{s,i}$. Nevertheless, its variation with respect to $x_{s,i}$
yields
\be
\sum_t C_{st} \bx_t = \sum_t C_{st} \,\partial_{\bw_t} \l_t \Big(\b\sum_q C_{tq} {\bf x}_q \Big) =  \sum_t C_{st} {\bf B}_t \Big(\b\sum_q C_{tq} {\bf x}_q \Big)\ .
\ee
Thus, if the coupling constant matrix ${C}_{st}$ is nonsingular, and therefore invertible, the above reproduces the equilibrium equations \eqn{sa}, including
the constraints \eqn{cons}, without restricting the variations of $x_{i,s}$ on the constraint space.
If, however, $C_{st}$ is singular, the equilibrium free energy misses the equations corresponding to the zero modes of $C_{st}$ and adds spurious solutions.
We stress that the zero modes of a singular coupling matrix do not drop out of the system, since they still appear in the arguments of the characters $\chi_s$.

\no
The issue of whether the equilibrium free energy can reproduce the correct stability conditions is even subtler.
To address it without delving in the specific details of the ferromagnet, we consider a general situation in which there are $p$ physical variables
$x_i$ and $p$ auxiliary variables $w_i$, with an effective free energy $f(\{x_i,w_i\})$ (in the case of our magnet, $p=mN$). The saddle-point equations of
$f$ in $w_i$ determine the true free energy $F(\{ x_i \})$, and the saddle point of $F$ in $x_i$ determines the equilibrium configurations.

\no
Close to a solution of the saddle-point equations, the free energy $f$ will generically be quadratic in the deviations of $x_i$ and $w_i$ from the solution. Calling the
deviations again $x_i$ and $w_i$, the general form of $f$ near a saddle point is
\be
f = \sum_{i,j} \left(- \half A_{ij} x_i x_j - \half B_{ij} w_i w_j + C_{ij} x_i w_j \right)\ ,
\ee
with $A,B,C$ being $p$-dimensional square matrices of coefficients. The symmetric matrices $A$ and $B$ may have zero modes, 
but we take $C$ to be nonsingular so that variables $x_i$ and $w_i$ are fully coupled. In that case, a redefinition of $w_i$ can bring $C$ to the identity, and we take $C = \mathbb{1}$ from now on.
The saddle point equations
in a matrix-vector form are,
\be
x:\ w=Ax \ , \quad w:\ x=Bw\ .
\ee
Using $w$ from the first equation, the equilibrium equation for $x$ is
\be
(1 - BA) x = 0\ .
\label{ec}
\ee
This admits the solution $x=0$ but, for $1-BA$ singular, it also has other solutions. In this case, retaining higher order terms in $f$ is necessary.

\no
If $B$ is nonsingular, the $w$  saddle point equation can be solved for $w$. Plugging back into $f$ gives the off-equilibrium free energy
\be
F_o = \ha x^T  (B^{-1} - A) x\ .
\ee
The equilibrium condition from $F_o$ is $(B^{-1} -A) x =0$ which is \eqn{ec} upon multiplying with $B$. 
On the other hand, the saddle point equation for $x$ can be solved for
$w$, and plugging back into $f$ gives the equilibrium free energy
\be
F_e = \ha x^T (A - ABA) x\ ,
\ee
which gives as equilibrium equation $A(1-BA) x =0$. This is equivalent to \eqn{ec} only if $A$ is nonsingular, otherwise it misses some
of the equations and introduces spurious solutions.

\no
The stability of the solution $x=0$ is determined by the Hessian $H_o = B^{-1} - A$ of $F_o$. On the other hand, the stability condition derived
from $F_e$ is determined by the Hessian $H_e = A(1-BA)$. The two are related as $H_e = AB H_o$ and in general do not lead to the same
positivity and stability criteria.

\no
In the case that $A$ and $B$ have definite positivity properties (either positive definite or negative definite), then either $H_e$ or $-H_e$
will lead to the correct stability conditions, since the product of a matrix with a positive definite matrix does not change its positivity properties.
(Proof: for positive definite $A$, the square root matrices $A^{\pm1/2}$ are defined, and $AB = A^{1/2} (A^{1/2} B A^{-1/2}) A^{1/2}$. The matrix
in parenthesis has the same eigenvalues as $B$, and since
\be
x^T A^{1/2} \big(A^{1/2} B A^{-1/2}\big) A^{1/2} x = (A^{1/2} x)^T \big(A^{1/2} B A^{-1/2}\big) (A^{1/2} x)\ ,
\ee
for any $x$, conjugating it with the positive definite matrix $A^{1/2}$ does not change its positivity
properties.) So, if $A,B$ are both positive or both negative definite, $H_e$ leads to the same conditions as $H_o$, while if one is positive
and the other negative, $-H_e$ does. However, if $A$ and $B$ have no definite positivity properties, $H_e$ does not lead to the correct stability
conditions.

\no
In the case of the magnet, comparing with the free energy  \eqn{c4} we see that $A$ corresponds to $\bf C$, which is not necessarily positive definite, or even nonsingular, and $B$ to $\bf \Lambda$,
which has the constraint vectors ${\bf u}_s$ as zero modes. This means that the saddle equation for $w$ can be solved only if $x$ is in the range
of $\bf \Lambda$, that is, if $x$ satisfies the constraints, and it leaves $w$ undetermined up to terms proportional to the constraint vectors, while
the equilibrium free energy will give the full equilibrium equations of $x$ upon varying $x$ only if $\bf C$ is nonsingular, facts we have already
encountered. Further, the stability conditions obtained from the equilibrium free energy are not in general the correct ones, and are guaranteed
to be correct only if both $\bf C$ and $\bf \Lambda$ are positive definite.

\section{Proof of positivity of $\L_r$}

For $U$ a diagonal matrix
\begin{equation}
    U = \diag (e^{ w_1}, \ldots, e^{ w_N})\ ,
\end{equation}
the character in an irrep $r$ of $SU(N)$ becomes
\begin{equation}
    \chi_r(\bw) = \text{tr}_r \,\diag\{e^{w_i}\} = \sum_\mu e^{ \bq_\mu \cdot  \bw}\ ,
\end{equation}
where $ \bq_\mu = ( q_{\mu 1}, \ldots,  q_{\mu N})$, $\mu = 1,2,\dots$ are the eigenvalues of the diagonal operators
in irrep $r$, that is, the weight vectors of $r$.
Differentiating $\l_r = \ln \chi_r$ with respect to $w_i$
\be
    \L_{r,ij} = \frac{\partial^2 \l_r}{\partial  w_i \, \partial  w_j} 
    = \frac{\sum_\mu \,  q_{\mu i} \,  q_{\mu j}\, e^{ \bq_\mu \cdot  \bw}}{\chi_r}
       - \frac{\sum_\mu \,  q_{\mu i}\, e^{ \bq_\mu \cdot  \bw}}{\chi_r}
         \, \frac{\sum_\nu \,  q_{\nu j}\, e^{ \bq_\nu \cdot  \bw}}{\chi_r}\ .
\ee
So for any vector $\bv \neq 0$
\begin{equation}
  \bv^T {{\bf \Lambda}_r} \bv = \sum_{i,j} v_i \L_{r,ij} v_j
  = \frac{\sum_\mu \,  (\bq_{\mu}\bb \cdot \bv)^2\, e^{ \bq_\mu \cdot  \bw}}{\chi_r}
   - \left(\frac{\sum_\mu \,  \bq_{\mu}\bb\cdot \bv \, e^{ \bq_\mu \cdot  \bw}}{\chi_r} \right)^2
   \geqslant 0\ ,
\end{equation}
by the Cauchy-Schwartz inequality, since $e^{ \bq_\mu \cdot  \bw}$ and $\chi_r$ are positive. Therefore, ${\bf\Lambda}_r$ is a positive definite matrix.
Equality holds only if $ \bq_\mu \cdot \bv$ is the same for all $\mu$, which happens
if $v_i = c$ and reproduces the zero mode of ${\bf\Lambda}_r$.

\end{document}